\documentclass[onecolumn,secnumarabic,amssymb,aps,prd,showkeys]{revtex4-2}

\usepackage{algorithmic}

\makeatletter
\newcommand\ext@algorithm{loa}
\newcommand\fps@algorithm{tbp}
\newcommand\ftype@algorithm{4}
\newcounter{algorithm}
\newcommand\fname@algorithm{Algorithm}
\newcommand\fnum@algorithm{\fname@algorithm\nobreakspace\thealgorithm}
\newenvironment{algorithm}
  {\@float{algorithm}}
  {\end@float}
\makeatother

\usepackage{graphicx}
\usepackage{amsmath}
\usepackage{amssymb}
\usepackage{bm}
\usepackage{booktabs}
\usepackage{enumerate}
\usepackage{hyperref}

\graphicspath{{figs/}}

\newcommand{\DCtwo}{DC\textsuperscript{2}SBM}
\newcommand{\Poisson}{\mathrm{Poisson}}
\newcommand{\etatrue}{\eta^{\text{true}}}
\newcommand{\rhotrue}{\rho^{\text{true}}}
\newcommand{\etahat}{\hat{\eta}}
\newcommand{\rhohat}{\hat{\rho}}

\begin{document}

\renewcommand{\thesection}{\arabic{section}}
\renewcommand{\thesubsection}{\thesection.\arabic{subsection}}
\renewcommand{\thesubsubsection}{\thesubsection.\arabic{subsubsection}}

\title{Dual-Corrected Stochastic Block Model (\DCtwo): A Generalized Structure Detection Method for Attributed Networks}

\author{Zhenhai Chang}
\email{changzhenhai2012@tsnu.edu.cn}
\affiliation{School of Mathematics and Statistics, Tianshui Normal University, Tianshui, Gansu, China}

\author{Shuo Wang}
\affiliation{School of Mathematics and Statistics, Tianshui Normal University, Tianshui, Gansu, China}

\author{Yumeng Liu}
\affiliation{School of Mathematics and Statistics, Tianshui Normal University, Tianshui, Gansu, China}

\author{Wei Liu}
\email{liuwei2007@tsnu.edu.cn}
\affiliation{School of Mathematics and Statistics, Tianshui Normal University, Tianshui, Gansu, China}

\begin{abstract}
Community detection in attributed networks is one of the core tasks of complex network analysis. The stochastic block model (SBM) and its variants are able to detect generalized network structures such as assortative, disassortative and mixed structures. However, existing generative methods either ignore node heterogeneity, or characterize node activity in the topological and attribute dimensions by one and the same factor, making it difficult to correctly model node activity when the two kinds of activity are decoupled. In this paper, we propose the Dual-Corrected Stochastic Block Model (\DCtwo), which introduces independent correction factors---a \emph{topological degree factor} and an \emph{attribute abundance factor}---into the link-generation and attribute-generation channels, respectively, thereby effectively decoupling node activity from membership inference, and which further introduces a global scaling parameter to resolve the mismatch between the normalization constraints and the network scale. The model is inferred by the expectation--maximization (EM) algorithm, and the newly introduced factors cancel out completely in the posterior computation, so that no additional computational cost is incurred. Experiments on synthetic and real attributed networks show that \DCtwo\ outperforms state-of-the-art methods in community detection accuracy, with an especially pronounced advantage when the topological signal is weak. The two correction factors can be recovered independently from their respective channels without cross-talk. Analytical as well as controlled experiments further demonstrate that modeling with a single shared factor fails when the two kinds of activity are decoupled---a regime that is prevalent in real networks---thereby validating the necessity of dual independent correction.
\end{abstract}


\maketitle

\section{Introduction}

Complex systems are ubiquitous in many domains of the real world, such as social, biological and information networks, and community detection, as one of the core tasks of network analysis, aims to reveal the latent functional modules and organizational regularities of a network \cite{GirvanNewman2002,Fortunato2010}. Traditional community detection methods rely mainly on the network topology and partition communities according to the connections between nodes \cite{ShiMalik2000,Yang2014}. However, as network data become increasingly abundant, nodes themselves often carry a large amount of attribute information, which provides important complementary information for community detection \cite{Shiga2007,Jia2017}.

Fusing network topology and node attributes for community detection has become an active research area in recent years \cite{Bothorel2015,Chunaev2020}. The homophily principle states that nodes with similar attributes tend to connect to each other, which provides an intuitive theoretical basis for exploiting attribute information \cite{Shalizi2011}. In real networks, however, nodes not only exhibit a power-law distribution in their number of connections (i.e., their degree), but also often differ greatly in attribute abundance (i.e., the number of attributes possessed by a node). Without correction, a model readily groups highly active nodes---whether in degree or in attribute abundance---into the same community, thereby masking the true community structure.

Within the framework of generative models, the stochastic block model (SBM) \cite{Holland1983} has attracted wide attention because it can characterize a variety of network structures, including assortative, disassortative and mixed structures \cite{NewmanPeixoto2015,Shen2011}. The degree-corrected stochastic block model (DCSBM) proposed by Karrer and Newman \cite{KarrerNewman2011} introduces a node-level degree correction factor into link generation, which effectively controls the influence of node activity on edge probability and substantially improves the fit to real networks. DCSBM, however, exploits only topological information and cannot incorporate node attributes. Subsequently, the PSB\_PG model proposed by Chang et al.\ \cite{Chang2019} brings node attributes into the generative process and achieves unified modeling of topology and attributes through a shared soft membership matrix, while retaining the ability of the SBM to detect generalized structures. Yet PSB\_PG does not correct node heterogeneity in either the topological or the attribute generation process---it implicitly assumes that every node has the same ``baseline capacity'' for generating links or attributes---so that nodes with high degree or high attribute counts are easily merged into a single community, again masking the true structure. On this basis, Zheng et al.\ proposed the DPSB\_PG model \cite{Zheng2020}, which introduces the observed node degree $x_i$ as a correction factor simultaneously into the two generative channels in order to fit the power-law degree distribution of real networks. DPSB\_PG, however, \emph{shares} one and the same degree factor across the two channels, thereby implicitly assuming that a node's topological activity and attribute abundance are perfectly consistent; it therefore models poorly when the node degree and the attribute count differ substantially. Wang et al.\ \cite{Wang2025} extended the idea of correcting by a single node degree to a weighted combination of several topological features, but likewise assumes that every node has the same capacity for generating links or attributes. Moreover, existing attribute-aware generative models such as CESNA \cite{Yang2013} and NEMBP \cite{He2017} also fail to account simultaneously for the twofold heterogeneity of topological degree and attribute abundance.

Based on the above observations, this paper proposes the Dual-Corrected Stochastic Block Model (\DCtwo). The main contributions are as follows.

\begin{enumerate}[(1)]
	\item To overcome the limitation of existing generative models, which characterize the two kinds of activity by a single factor, mutually independent correction factors are introduced into the topological and the attribute channel, respectively, achieving independent modeling and complete decoupling of the twofold heterogeneity, with both factors updated adaptively by the EM algorithm rather than supplied as fixed inputs. A global scaling parameter is introduced to resolve the mismatch between the normalization constraints and the network scale. The model reduces to DPSB\_PG \cite{Zheng2020} or PSB\_PG \cite{Chang2019} as special cases, which confers a certain generality.
	
	\item We derive the EM parameter-inference procedure of the model and provide a vectorized implementation together with a complexity analysis. The per-iteration complexity is linear in the number of edges and the attribute scale, which guarantees the feasibility of the algorithm on moderately large
	networks.
	
	\item Experiments on synthetic and real attributed networks verify the superiority of \DCtwo\ in generalized structure detection. Furthermore, the dual-correction mechanism is examined systematically from three aspects---factor recoverability, decoupling and necessity.
\end{enumerate}

\section{Related work}

\subsection{Community detection based on stochastic block models}

The stochastic block model (SBM) \cite{Holland1983} is a classical paradigm of generative community detection. Its core assumption is that the probability of an edge between two nodes depends only on the communities to which they belong and is not affected by individual differences between nodes. The block-structure matrix $\Theta$ of the model can characterize a variety of network types: when the diagonal elements are large the model corresponds to an assortative (traditional community) structure, when the off-diagonal elements are large it corresponds to a disassortative structure, and other cases correspond to mixed structures \cite{NewmanPeixoto2015,Shen2011}. The mixed-membership model (NMM) proposed by Newman and Leicht \cite{NewmanLeicht2007} further introduced the notion of soft membership, allowing a node to be distributed over several communities with a certain probability.

The standard SBM, however, does not account for the power-law degree distribution of real networks. The degree-corrected stochastic block model (DCSBM) proposed by Karrer and Newman \cite{KarrerNewman2011} solved this problem by introducing the node degree parameter $\theta_i$, modeling the edge probability as $p_{ij}\propto\theta_i\theta_j\omega_{z_i,z_j}$. This effectively decouples the contributions of individual node activity and of community-level connection strength to the edge probability, and avoids the bias of the standard SBM whereby high-degree nodes dominate community inference. The model fits real networks markedly better than the standard SBM and has become the basis of many subsequent extensions. On this basis, the general stochastic block (GSB) model proposed by Shen et al.\ \cite{Shen2011} further broadened the applicability of the SBM, being able to detect generalized structures in directed networks and to measure node centrality.

\subsection{Generative models incorporating node attributes}

To incorporate attribute information, researchers have proposed a variety of generative models. From the perspective of whether a node-level correction factor is introduced, these models fall into two categories.

The first category comprises models that do not consider node correction factors. For instance, the CESNA model proposed by Yang et al.\ \cite{Yang2013} adopts a hard-membership assumption and incorporates attributes into the likelihood through logistic regression, but it mainly targets traditional community structures and does not consider degree correction. The PSB\_PG model proposed by Chang et al.\ \cite{Chang2019} models links and attributes separately by Poisson distributions, unifies the two kinds of information through a shared soft membership matrix $D$, and uses the EM algorithm to guarantee convergence. This model can detect generalized structures and is semantically interpretable, being able to extract relevant attribute keywords for each community on real datasets. PSB\_PG, however, introduces no node-level correction factor into either the link or the attribute generation process; it implicitly assumes that all nodes share the same ``baseline activity'', and performs poorly when the node degree and the attribute count differs greatly across nodes.

The second category comprises models that consider or borrow the idea of degree
correction. This category can be further divided, according to the source of the
correction factor and the channel on which it acts, into three subclasses:
correction of the topological channel only, shared correction of both channels by the same factor, and topological correction derived from attributes.

\textbf{(1) Correction of the topological channel only.} Yang et al.\ proposed
the PCL\_DC model \cite{Yang2009}, which introduces a ``popularity'' variable
$b_i$ to measure the likelihood of a node being connected by other nodes, thereby
separating the individual connection propensity of a node from the community
structure. In the subsequent PPL\_DC model \cite{Yang2010}, popularity $a_i$ and
productivity $b_i$ are further introduced; through the introduction of these two
independent parameters, the model captures the difference between the two roles
of ``being connected'' and ``actively connecting'' in directed networks, and thus
describes the network formation mechanism more precisely. The PPSB\_DC model
proposed by Chai et al.\ \cite{Chai2013} inherits both the block-structure
assumption of the SBM and the degree-distribution modeling capability of PPL, and
can detect generalized structures. The NEMBP model proposed by He et al.\
\cite{He2017} combines the degree-corrected stochastic block model with a
multinomial distribution and solves it by nested EM and belief propagation,
offering good semantic interpretability. A difference is that the node degree
correction factors in the first three models evolve adaptively with the
iterations of the algorithm, whereas NEMBP fixes the node degree correction
factors in advance and keeps them constant throughout the iterations. However,
PCL\_DC, PPL\_DC, PPSB\_DC and NEMBP all consider only the topological
heterogeneity of nodes and neglect their attribute heterogeneity.

\textbf{(2) Shared correction of both channels by the same factor.} On
the basis of PSB\_PG, Zheng et al.\ proposed the DPSB\_PG model \cite{Zheng2020},
which introduces the observed node degree $x_i$ as a correction factor
simultaneously into the two generative channels of links and attributes.
Specifically, the link expectation is
$\hat A_{ij} = \sum_{r,s} x_i d_{ir}\theta_{rs}x_j d_{js}$ and the attribute
expectation is $\hat W_{ik} = \sum_r x_i d_{ir}\phi_{rk}$, where $x_i$ is the
observed degree of node $i$, computed once at initialization and then supplied as
a fixed input participating in the EM iterations. This strategy fits the
power-law degree distribution of real networks to some extent. DPSB\_PG,
however, shares the same degree factor $x_i$ across the two channels of topology
and attributes, implicitly assuming that a node's topological activity and
attribute abundance are perfectly consistent; it therefore models poorly when
the node degree and the attribute count differ greatly.

Recently, Wang et al.\ \cite{Wang2025} generalized the single-node-degree
correction of DPSB\_PG to a weighted combination of several topological features,
introducing the degree, betweenness centrality and clustering coefficient of a
node and controlling their contributions through the parameters
$\alpha,\beta,\eta$ to construct a composite influence
$\delta_i = \alpha y_i + \beta b_i + \eta c_i$ as the node correction factor.
This model (BCSBM) likewise models links and attributes by Poisson distributions
and solves them with the EM algorithm. As with DPSB\_PG, however, the
composite-influence correction factor is computed once at initialization and is
introduced simultaneously into the topological and the attribute information,
implicitly assuming that a node's topological activity and attribute abundance
are perfectly consistent.

\textbf{(3) Topological correction derived from attributes.} CRSBM
\cite{Ren2023} adopts a route different from all the above methods: it does not
introduce an independent node correction factor, but instead converts the
attribute information into a node popularity factor and embeds it into the SBM;
its topological correction is therefore derived from the attributes, while the
attribute channel itself is not modeled.

Beyond the above generative models, non-generative routes to community detection
on attributed networks also exist, including spectral methods and non-negative
matrix factorization, as well as the rapidly developing graph neural network
(GNN) based deep models \cite{Chunaev2020}. Such methods are trained end to end
with a reconstruction error or a contrastive-learning loss and often perform
well on strongly assortative networks; most of them, however, implicitly assume
homophily and struggle to characterize generalized structures such as
disassortative, bipartite and mixed structures, and they provide neither
parameter-level interpretation of a generative mechanism nor statistical
inference properties. The present work focuses on the generative stochastic block
model route, with the goal of solving the modeling problem of twofold node
heterogeneity while preserving statistical interpretability and the ability to
detect generalized structures.

In view of the above issues, this paper proposes the Dual-Corrected Stochastic
Block Model (\DCtwo), whose key difference from the preceding methods lies in the
independence and the update scheme of the factors; a detailed comparison is given
in Table~\ref{tbl:methods}.

\begin{table}[htbp]
\caption{Comparison of the methods. The symbol ``---'' indicates that the
corresponding property is absent. Only representative generative models for
generalized structure detection are listed; PCL\_DC \cite{Yang2009} and
PPL\_DC \cite{Yang2010}, which mainly target traditional community structure in
directed networks and whose degree-correction idea has been inherited and
developed by PPSB\_DC \cite{Chai2013}, are therefore not listed
separately.}\label{tbl:methods}
\begin{tabular}{@{}lcccc@{}}
\toprule
Method & Topological correction & Attribute correction & Factor independence & Update scheme\\
\midrule
PSB\_PG \cite{Chang2019}              & $\times$ & $\times$ & --- & ---\\
NEMBP \cite{He2017}                   & $\checkmark$ ($d_i$) & $\times$ & --- & Fixed prior input\\
PPSB\_DC \cite{Chai2013}              & $\checkmark$ ($a_i,b_i$) & $\times$ & --- & Adaptive update\\
DPSB\_PG \cite{Zheng2020}             & $\checkmark$ (observed degree $x_i$) & $\checkmark$ (same $x_i$) & Shared & Fixed prior input\\
BCSBM \cite{Wang2025}                 & $\checkmark$ (multi-indicator combination $\delta_i$) & $\checkmark$ (same $\delta_i$) & Shared & Fixed prior input\\
CRSBM \cite{Ren2023}                  & $\checkmark$ ($f(\alpha_{ir})$) & $\times$ & --- & Adaptive update\\
\textbf{\DCtwo\ (this paper)}         & $\checkmark$ ($\rho_i$) & $\checkmark$ ($\eta_i$) & \textbf{Independent} & \textbf{Adaptive update}\\
\bottomrule
\end{tabular}
\end{table}

\section{Dual-Corrected Stochastic Block Model (\DCtwo)}

This section introduces the notation of the proposed model and its generative process.

\subsection{Notation and definitions}

Let $G=(V,E,W)$ be an attributed network, where $V=\{1,2,\ldots,n\}$ is the set
of $n$ nodes and $E$ is the edge set. The adjacency matrix
$A=(A_{ij})_{n\times n}$ is a symmetric integer matrix that allows multi-edges
and contains no self-loops, and the total number of edges is
$m=\frac{1}{2}\sum_{i,j}A_{ij}$. The attribute matrix $W=(W_{ik})_{n\times K}$ is
a count matrix, where $K$ is the attribute dimension and $W_{ik}$ denotes the
frequency with which node $i$ possesses attribute $k$; $w_i=\sum_k W_{ik}$ is the
attribute abundance of node $i$ and $\bar{w}=\frac{1}{n}\sum_i w_i$. To
distinguish the heterogeneity of a node along the two dimensions of topological
linking and attribute generation, two correction factors are introduced---the
\textbf{topological degree factor} $\rho_i$ and the \textbf{attribute abundance
factor} $\eta_i$.

\begin{itemize}
\item \textbf{Topological degree factor} $\rho_i > 0$: controls the ``activity''
of node $i$ in link generation. A high-degree node has a larger $\rho_i$, so that
its expected number of edges is amplified, which prevents the model from
mistaking ``high degree'' itself for community structure.
\item \textbf{Attribute abundance factor} $\eta_i > 0$: controls the ``activity''
of node $i$ in attribute generation. A node with a high attribute count (e.g., a
user with many publications or many tags) has a larger $\eta_i$, so that its
expected number of attributes is amplified, which prevents the model from
mistaking ``many attributes'' itself for a signal of community membership.
\end{itemize}

The two factors are mutually independent and separately eliminate the
heterogeneity interference in the topological and the attribute modalities,
achieving independent modeling and complete decoupling of the twofold node
activity. The main symbols involved in the model are summarized in
Table~\ref{tbl:notation}.

\begin{table}[htbp]
\caption{Notation of the \DCtwo\ model.}\label{tbl:notation}
\begin{tabular}{@{}lp{0.74\linewidth}@{}}
\toprule
Symbol & Description\\
\midrule
$n$ & total number of nodes\\
$m$ & total number of edges\\
$K$ & attribute dimension\\
$c$ & number of communities\\
$A=(A_{ij})_{n\times n}$ & adjacency matrix; $A_{ij}$ is the number of edges between nodes $i$ and $j$\\
$W=(W_{ik})_{n\times K}$ & attribute matrix; $W_{ik}$ is the number of times node $i$ possesses the $k$-th attribute\\
$w_i=\sum_k W_{ik}$ & attribute abundance of node $i$\\
$g_i=\sum_j A_{ij}$ & degree of node $i$\\
$D=(d_{ir})_{n\times c}$ & soft membership matrix; $d_{ir}$ is the strength with which node $i$ belongs to community $r$\\
$\Theta=(\theta_{rs})_{c\times c}$ & inter-community connection strength matrix; $\theta_{rs}$ is the connection probability between communities $r$ and $s$\\
$\Phi=(\phi_{rk})_{c\times K}$ & community--attribute probability matrix; $\phi_{rk}$ is the probability that community $r$ possesses the $k$-th attribute\\
$\rho=(\rho_i)$ & topological degree factor, absorbing the heterogeneity of nodes in edge generation\\
$\eta=(\eta_i)$ & attribute abundance factor, absorbing the heterogeneity of nodes in attribute generation\\
$\omega$ & global scaling parameter, controlling the total number of edges of the network\\
\bottomrule
\end{tabular}
\end{table}

The model parameters satisfy the following identifiability constraints:
\begin{equation}
\sum_{i=1}^n \rho_i d_{ir}=1,\quad \forall r,\qquad \sum_{r,s}\theta_{rs}=1,\qquad \sum_i\eta_i=n,\qquad \sum_k\phi_{rk}=1.
\label{eq:constraints}
\end{equation}

\subsection{Generative process}

Unlike DCSBM \cite{KarrerNewman2011}, which allows self-loops to simplify the
derivation, we consider an undirected network without self-loops and define the
edge generation process only for unordered node pairs $(i<j)$. Self-loops are
extremely rare in real sparse networks, and neglecting them affects the expected
node degree only by a relative amount of $O(1/n)$, so this setting does not
affect the core results of community detection. \DCtwo\ assumes that both the
edges of the network and the node attributes are generated by Poisson
distributions, and that the two are conditionally independent given the soft
membership matrix $D$.

\textbf{Edge generation.} In the classical SBM, the edges between nodes are
controlled entirely by the community-structure parameter
$\Theta=(\theta_{rs})_{c\times c}$. To relax this setting, PSB\_PG
\cite{Chang2019} introduced the parameter $D=(d_{ir})_{n\times c}$, using
$d_{ir}$ to characterize the propensity of node $i$ towards community $r$, so
that the propensity for an edge between nodes $(i,j)$ in the community pair
$(r,s)$ is $d_{ir}\theta_{rs}d_{js}$. Drawing on the node degree correction of
DCSBM \cite{KarrerNewman2011}, we introduce a node degree correction factor
$\rho_i>0$ to further separate the influence of large degree differences among
nodes on the community structure $\Theta$, so that the propensity for an edge
between nodes $(i,j)$ in the community pair $(r,s)$ becomes
$\rho_i\rho_j d_{ir}\theta_{rs}d_{js}$. Summing over all community pairs, the
expected number of edges between nodes $i$ and $j$ is
\begin{equation}
\hat A_{ij} = \omega\,\rho_i\rho_j\sum_{r,s} d_{ir}\theta_{rs}d_{js} = \omega\,\rho_i\rho_j B_{ij},\quad B_{ij}=\sum_{r,s}d_{ir}\theta_{rs}d_{js}.
\label{eq:Ahat}
\end{equation}
Here $B_{ij}$ reflects the baseline connection strength of nodes $i$ and $j$ at
the community level; $\rho_i\rho_j$ is the topological degree correction term,
which gives high-degree nodes a higher expected number of edges and prevents
high-degree nodes from being mistaken for community structure; and $\omega$ is
the global scaling parameter, explained below.

\textbf{Global scaling.} To guarantee the identifiability of the parameters, the
constraint $\sum_{i=1}^n \rho_i d_{ir} = 1, \ \forall r$, is imposed. This
constraint gives $\theta_{rs}$ a direct physical meaning, namely the expected
total number of edges between communities $r$ and $s$ (since
$\sum_{i,j} \rho_i d_{ir} \theta_{rs} \rho_j d_{js} = \theta_{rs}$). Without
$\omega$, the constraints $\sum_i\rho_i d_{ir}=1$ and $\sum_{r,s}\theta_{rs}=1$
would imply $\sum_{i,j}\rho_i\rho_j B_{ij}=1$, which severely mismatches the true
total number of edges $2m$. Hence $\omega>0$ is introduced so that
$\sum_{i,j}\hat A_{ij}=\omega\approx 2m$, allowing the model to adapt to networks
of arbitrary size, with the overall edge count absorbed entirely by $\omega$.

Based on the above parameters, we assume that the edges follow a Poisson
distribution, that is,
$P(A\mid \omega, \rho, D, \Theta) \sim \Poisson(\hat A)$.

\textbf{Attribute generation.} Under the above parameters, the probability that
node $i$ generates attribute $k$ via community $r$ is given by $d_{ir}\phi_{rk}$.
To avoid the influence of ``many attributes'' itself on community membership, we
introduce the attribute abundance factor $\eta_i>0$, which controls the
``activity'' of node $i$ in attribute generation and separates the impact of node
attribute differences on the parameter $\phi_{rk}$. We assume that the number of
times $W_{ik}$ that node $i$ possesses the $k$-th attribute follows a Poisson
distribution with expectation
\begin{equation}
\hat W_{ik} = \eta_i\sum_r d_{ir}\phi_{rk} = \eta_i P_{ik},\quad P_{ik}=\sum_r d_{ir}\phi_{rk}.
\label{eq:What}
\end{equation}
Here $P_{ik}$ reflects the baseline attribute-generation probability of node $i$
at the community level, and $\eta_i$ is the attribute abundance correction term,
which gives nodes with high attribute counts a higher expected number of
attributes. It is independent of $\rho_i$ in the edge channel, allowing a node to
be highly active topologically yet sparse in attributes (or vice versa), thereby
decoupling the two kinds of heterogeneity. The constraint $\sum_i\eta_i=n$,
i.e., the mean attribute abundance factor over all nodes is $1$, preserves the
meaning of $\phi_{rk}$ as ``the baseline probability of attribute $k$ in
community $r$''. The Poisson process of attribute generation automatically
matches the observed counts, so no additional total-amount constraint is needed.

The soft membership matrix $D$ determines the latent position of the nodes.
Assuming that, given a shared latent node position, the topological and the
attribute information of a node are independent, the joint likelihood of the
observations $(A,W)$ satisfies $P(A,W\mid D)=P(A\mid D)P(W\mid D)$, where all
parameters other than $D$ are suppressed. Omitting irrelevant constant terms, the
joint log-likelihood of the observations $(A,W)$ is
\begin{equation}
\begin{aligned}
\mathcal{L} = \sum_{i<j}\left[A_{ij}\log(\omega\rho_i\rho_j B_{ij}) - \omega\rho_i\rho_j B_{ij}\right]
+ \sum_{i,k}\left[W_{ik}\log(\eta_i P_{ik}) - \eta_i P_{ik}\right].
\end{aligned}
\label{eq:loglik}
\end{equation}

\textbf{Relationship to existing models.} By different settings of the two
correction factors, \DCtwo\ subsumes several existing models as special cases:
(i) setting $\rho_i\equiv1$ and $\eta_i\equiv1$, the model reduces, under the
reparameterization $\omega\theta_{rs}\to\theta_{rs}$, to PSB\_PG
\cite{Chang2019}, where the constraint $\sum_i\rho_i d_{ir}=1$ corresponds to the
normalization convention of PSB\_PG for $D$; (ii) setting only $\eta_i\equiv1$,
the attribute channel no longer absorbs node heterogeneity and the model reduces
to a ``topology-corrected, attribute-uncorrected'' variant whose topological
channel coincides with the soft-membership version of DCSBM
\cite{KarrerNewman2011}; (iii) setting $\rho_i=\eta_i=x_i$ and fixing it at the
observed degree at initialization (without further updates during the
iterations), the model reduces to the shared-correction modeling of the
DPSB\_PG \cite{Zheng2020} type, with BCSBM \cite{Wang2025} corresponding to the
same setting fixed to the composite indicator $\delta_i$; and (iv) if
$\rho_i\equiv\eta_i$ is kept but still updated adaptively by the EM algorithm,
the result corresponds to the shared variant examined in
Section~\ref{sec:necessity}. Therefore, the difference between the present model
and existing methods can be uniformly reduced to two points: whether the two
factors are independent, and whether the factors are updated adaptively
(Table~\ref{tbl:methods}).

\section{Model inference and algorithm design}

In Eq.~\eqref{eq:loglik}, since a summation appears inside the logarithm,
the parameters cannot be solved directly. We therefore adopt the EM algorithm and
optimize iteratively by introducing latent variables. Below, we first derive the
update formulas of all parameters in the E-step and the M-step separately, and
then present an efficient vectorized implementation, the algorithmic workflow and
the computational complexity. To keep the main text concise, the details of some
derivations are deferred to the appendices.

\subsection{E-step: posterior computation of the latent variables}

We introduce the latent variable $Z_{ij}^{rs}\in\{0,1,2,\dots\}$, denoting the
number of edges between the node pair $(i,j)$ that are generated by the community
pair $(r,s)$, and the latent variable $Y_{ik}^{r}\in\{0,1,2,\dots\}$, denoting the
number of times the $k$-th attribute of node $i$ is generated by community $r$.
They satisfy the constraints
\begin{equation*}
\sum_{r,s} Z_{ij}^{rs}=A_{ij},\qquad \sum_{r} Y_{ik}^{r}=W_{ik}.
\end{equation*}

Assuming that each community pair $(r,s)$ independently generates $Z_{ij}^{rs}$
edges following a Poisson distribution, the posterior probability of the edge
latent variable $Z_{ij}^{rs}$ under the constraint
$\sum_{r,s} Z_{ij}^{rs}=A_{ij}$ is given by (the detailed derivation is in Appendix~\ref{app:posterior}):
\begin{equation}
q_{ij}^{rs} = \frac{\omega\rho_i\rho_j d_{ir}\theta_{rs}d_{js}}
                {\sum_{r',s'}\omega\rho_i\rho_j d_{ir'}\theta_{r's'}d_{js'}}
            = \frac{d_{ir}\theta_{rs}d_{js}}{B_{ij}}.
\label{eq:qij}
\end{equation}

Similarly, assuming that each community $r$ independently generates $Y_{ik}^{r}$
attributes following a Poisson distribution, the posterior probability of the
attribute latent variable $Y_{ik}^{r}$ is obtained as follows (the detailed
derivation can be found in Appendix~\ref{app:posterior}):
\begin{equation}
\gamma_{ik}^r = \frac{\eta_i d_{ir}\phi_{rk}}{\sum_{r'}\eta_i d_{ir'}\phi_{r'k}}
            = \frac{d_{ir}\phi_{rk}}{P_{ik}}.
\label{eq:gamma}
\end{equation}

It should be noted that the global scaling parameter $\omega$ and the dual-correction
factors $\rho_i$ and $\eta_i$ all cancel out in the E-step, so the posterior computation incurs no additional cost from the parameters newly introduced by the model.

\subsection{M-step: parameter updates}

With the posterior expectations from the E-step, the expected complete-data
log-likelihood
$\mathbb{E}[\log p(A, W, Z, Y) \mid A, W, \text{other parameters}]$ is

\begin{equation}
\begin{aligned}
\bar{\mathcal L}
&=\sum_{i<j}\sum_{r,s}A_{ij}q_{ij}^{rs}
   \log\big(\omega\rho_i\rho_j d_{ir}\theta_{rs}d_{js}\big)
 -\sum_{i<j}\omega\rho_i\rho_j\sum_{r,s}d_{ir}\theta_{rs}d_{js}\\
&\quad +\sum_{i,k}\sum_r W_{ik}\gamma_{ik}^r
   \log\big(\eta_i d_{ir}\phi_{rk}\big) -\sum_{i,k}\eta_i\sum_r d_{ir}\phi_{rk}.
\end{aligned}
\label{eq:Q}
\end{equation}

In the M-step, the parameters are updated by maximizing $\bar{\mathcal L}$. Since the
parameters in the logarithmic terms are separable, each parameter can be solved
individually. Below we give, one by one, the closed-form or the univariate
numerical solution for each parameter; some detailed derivations are deferred to
Appendix~\ref{app:bisect}.

\subsubsection{Update of the global scaling $\omega$}

At iteration $t$, extracting the terms of Eq.~\eqref{eq:Q} involving $\omega$ and setting the derivative to zero gives

\begin{equation}
\omega^{(t+1)} = \frac{\sum_{i<j} A_{ij}}{\sum_{i<j} \rho_i^{(t)}\rho_j^{(t)} B_{ij}^{(t)}} = \frac{m}{\frac{1}{2}\sum_{r,s}\theta_{rs}^{(t)} U_r^{(t)} U_s^{(t)}} \approx 2m,
\label{eq:omega}
\end{equation}
where $U_r = \sum_i \rho_i d_{ir}$. Using the constraints $\sum_i\rho_i d_{ir}=1$
and $\sum_{r,s}\theta_{rs}=1$, the denominator is approximately $1/2$, so that on
large sparse graphs $\omega$ is approximately twice the total number of edges. In
practice the exact summation is used to guarantee the self-consistency of the
iterations.

\subsubsection{Update of the topological degree factor $\rho$}

Extracting the terms of $\bar{\mathcal L}$ involving $\rho_i$ and differentiating,
the first-order condition gives the unconstrained closed-form solution
\begin{equation}
\rho_i^{(t+1)} = \frac{g_i}{\omega^{(t+1)} \sum_{j \neq i} \rho_j^{(t)} B_{ij}^{(t)}},
\label{eq:rho_uncon}
\end{equation}
where $g_i = \sum_{j \neq i} A_{ij}$ is the observed degree of node $i$. The
denominator $\hat g_i = \omega \sum_{j \neq i} \rho_j B_{ij}$ is exactly the
expected degree of node $i$ under the current parameters. Using the
identifiability constraint $U_r = \sum_i \rho_i d_{ir} = 1$ and neglecting the
self-loop term in the large-graph sparse limit, this expected degree can be
simplified as
\begin{equation}
\begin{aligned}
\hat g_i
&= \omega \sum_{j \neq i} \rho_j \sum_{r,s} d_{ir} \theta_{rs} d_{js} = \omega \sum_{r,s} d_{ir} \theta_{rs} \Big( \sum_{j \neq i} \rho_j d_{js} \Big) \\
&\approx \omega \sum_{r,s} d_{ir} \theta_{rs} U_s
= \omega \sum_{s=1}^c d_{is} \sum_{r=1}^c \theta_{rs}.
\end{aligned}
\label{eq:ghat}
\end{equation}
Substituting Eq.~\eqref{eq:ghat} into Eq.~\eqref{eq:rho_uncon} gives the update
formula
\begin{equation}
\rho_i^{(t+1)} = \frac{g_i}{\omega^{(t+1)} \sum_{s=1}^c d_{is}^{(t)} \sum_{r=1}^c \theta_{rs}^{(t)}}.
\label{eq:rho}
\end{equation}

As the above derivation shows, $\rho_i$ is precisely the ratio of the observed
degree of node $i$ to its model-expected degree, which is fully consistent with
the construction of the classical degree-corrected model (DCSBM). Note that the
simplification here relies on the approximation $U_r\approx1$; in practice the
unconstrained update of $\rho$ makes $U_r$ deviate slightly from $1$ temporarily,
but the subsequent column-normalization projection of the membership matrix $D$
(see Section~\ref{sec:Dupdate}) immediately restores the constraint, so that no explicit rescaling of $\rho$ is required, which guarantees the convergence of the
whole EM procedure.

\subsubsection{Update of the attribute abundance factor $\eta$}

Under the constraint $\sum_i \eta_i = n$, maximizing $\bar{\mathcal L}$ gives
\begin{equation}
\eta_i = \frac{w_i}{\sum_r d_{ir} + \nu},
\label{eq:eta}
\end{equation}
where the Lagrange multiplier $\nu$ is determined by the equation
$\sum_{i} w_i / (\sum_r d_{ir} + \nu) = n$, which can be solved rapidly by
bisection; the detailed solution procedure is given in
Appendix~\ref{app:bisect}. If the variational approximation
$\sum_r d_{ir}\approx 1$ is adopted (which is reasonable when the cluster
structure is clear), the expression simplifies to
$\eta_i \approx w_i / \bar{w}$, namely the ratio of the total number of attributes
of a node to the network average.

\subsubsection{Update of the inter-community connection strength $\Theta$}

Define
\begin{equation}
F_{rs}^{\text{raw}} = \sum_{(i,j)\in E} \frac{d_{ir} d_{js}}{B_{ij}}.
\label{eq:Fraw}
\end{equation}
Then, using $q_{ij}^{rs} = d_{ir}\theta_{rs}d_{js}/B_{ij}$, the update formula
under the constraint $\sum_{r,s}\theta_{rs}=1$ is
\begin{equation}
\theta_{rs}^{(t+1)} = \frac{\theta_{rs}^{(t)} F_{rs}^{\text{raw}}}{\sum_{p,q}\theta_{pq}^{(t)} F_{pq}^{\text{raw}}}.
\label{eq:Theta}
\end{equation}
Written in terms of the adjacency matrix, this is equivalent to
$\theta_{rs} = \frac{1}{m}\sum_{i<j} A_{ij} q_{ij}^{rs}$, which has the intuitive
meaning of ``the proportion of edges between communities''.

\subsubsection{Update of the community--attribute probability $\Phi$}

Similarly, under the constraint $\sum_k \phi_{rk}=1$, maximizing
$\bar{\mathcal L}$ gives
\begin{equation}
\phi_{rk}^{(t+1)} = \frac{\phi_{rk}^{(t)} T_{rk}}{\sum_{k'}\phi_{rk'}^{(t)} T_{rk'}},
\quad T_{rk} = \sum_{i} W_{ik} \frac{d_{ir}}{P_{ik}}.
\label{eq:Phi}
\end{equation}
This expression keeps each row of $\Phi$ normalized.

\subsubsection{Update of the soft membership matrix $D$}
\label{sec:Dupdate}

The update of the membership matrix $D$ involves information from both the
topological and the attribute side. We first define the node--community composite
evidence matrix $\alpha \in \mathbb{R}^{n\times c}$, whose elements are the sum of
the topological evidence and the attribute evidence:
\begin{equation}
\alpha_{ir} = \sum_{j\in\mathcal{N}(i)} \frac{M_{jr}}{B_{ij}} + \sum_k W_{ik}\frac{\phi_{rk}}{P_{ik}},
\quad M = D\Theta.
\label{eq:alpha}
\end{equation}
Here $\mathcal{N}(i)$ denotes the set of neighbors of node $i$. Under the constraint $\sum_i \rho_i d_{ir}=1$, the Lagrange multiplier method gives the update formula
\begin{equation}
d_{ir}^{(t+1)} = \frac{d_{ir}^{(t)} \alpha_{ir}}{\rho_i(\omega C_r + \lambda_r) + \eta_i},
\qquad C_r = \sum_{s} \theta_{rs} U_s.
\label{eq:Dupdate}
\end{equation}
For each community $r$, the Lagrange multiplier $\lambda_r$ is the unique root of
the equation
\begin{equation}
\sum_{i=1}^n \frac{\rho_i d_{ir}^{(t)} \alpha_{ir}}{\rho_i(\omega C_r + \lambda_r) + \eta_i} = 1.
\label{eq:lambda}
\end{equation}
This equation is strictly monotonically decreasing on its domain and can be
solved efficiently by bisection; the details are given in
Appendix~\ref{app:bisect}. To eliminate numerical drift during the updates, a
column normalization is finally applied to each column $d_{*r}$:
\begin{equation}
d_{ir} \leftarrow \frac{d_{ir}}{\sum_j \rho_j d_{jr}}.
\label{eq:Dnorm}
\end{equation}

\subsection{Vectorized implementation and complexity analysis}

To improve computational efficiency, the EM updates above can be fully converted
into matrix/vector operations, avoiding explicit loops over node pairs $i<j$ or
over individual edges. Let $\odot$ and $\oslash$ denote element-wise
multiplication and division, respectively. At the beginning of each iteration,
three intermediate matrices are computed:
\begin{equation}
M = D\Theta \in \mathbb{R}^{n\times c},\quad
P = D\Phi \in \mathbb{R}^{n\times K},\quad
U = D^\top \rho \in \mathbb{R}^{c},
\label{eq:MPU}
\end{equation}
where $M_{ir}$ denotes the weighted connection potential of node $i$ through
community $r$, $P_{ik}$ is the baseline generation strength of node $i$ for
attribute $k$, and $U_r$ should be $1$ under the constraint. The statistic
$F^{\text{raw}}$ required by the E-step and the evidence matrix $\alpha$ can both
be obtained in one pass by vectorized accumulation over the edge list and matrix
multiplication, without explicitly storing the four-dimensional tensor
$q_{ij}^{rs}$ or the three-dimensional tensor $\gamma_{ik}^r$.

Table~\ref{tbl:complexity} summarizes all vectorized operations of the E-step and
the M-step together with their asymptotic complexity. The total complexity is
\begin{equation*}
O\big(m c^2 + n K c + n c\,T_{\text{bis}}\big),
\end{equation*}
where $T_{\text{bis}}$ is the number of bisection iterations (usually 40--80, see
Appendix~\ref{app:bisect}).

\begin{table}[htbp]
\caption{Vectorized EM algorithm and complexity.}\label{tbl:complexity}
{\footnotesize\setlength{\tabcolsep}{4pt}%
\begin{tabular}{@{}lp{0.33\linewidth}p{0.29\linewidth}l@{}}
\toprule
Step & Operation & Vectorized description & Complexity\\
\midrule
\textbf{Intermediate} & Compute $M = D\Theta$ & matrix multiplication & $O(nc^2)$\\
 & Compute $P = D\Phi$ & matrix multiplication & $O(nKc)$\\
 & Compute $U = D^\top\rho$ & matrix--vector product & $O(nc)$\\
\textbf{E-step} & Edge strength $B_{ij}$ & row-wise multiply--add indexed by edge & $O(mc)$\\
 & Accumulate $F^{\text{raw}}$ and symmetrize & outer-product accumulation by edge & $O(mc^2)$\\
 & Topological evidence $\alpha^{\text{(link)}}$ & scatter by edge and group-sum & $O(mc)$\\
 & Attribute evidence $\alpha^{\text{(attr)}} = (W\oslash P)\Phi^\top$ & matrix multiplication & $O(nKc)$\\
 & Composite evidence $\alpha = \alpha^{\text{(link)}}+\alpha^{\text{(attr)}}$ & matrix addition & $O(nc)$\\
\textbf{M-step} & Update $\omega$ & simplify via $U^\top\Theta U$ and the diagonal term & $O(nc^2)$\\
 & Update $\rho = \mathbf{g} \oslash (\omega\cdot D(\Theta U))$ & matrix--vector product and element-wise division & $O(nc^2)$\\
 & Update $\eta$ & bisection (Appendix~\ref{app:bisect}) & $O(n T_{\text{bis}})$\\
 & Update $\Theta \propto \Theta\odot F^{\text{raw}}$ & element-wise multiplication and normalization & $O(c^2)$\\
 & Update $\Phi \propto \Phi\odot (D^\top (W\oslash P))$ & matrix multiplication and element-wise multiplication & $O(nKc)$\\
 & For each $r$, find $\lambda_r$ by bisection and update $\mathbf{d}_{*r}$ & column-wise update and bisection & $O(nc\,T_{\text{bis}})$\\
 & Column-normalize $D$ & element-wise division & $O(nc)$\\
\bottomrule
\end{tabular}}
\end{table}

\subsection{Algorithmic workflow}

In summary, the inference algorithm of \DCtwo\ consists of two parts: the main EM
loop (Algorithm~\ref{alg:em}) and the bisection subroutine for solving the
Lagrange multipliers (Algorithm~\ref{alg:bisect}). The pseudocode is summarized
below.

\begin{algorithm}[!htbp]
\caption{EM algorithm for \DCtwo}\label{alg:em}
\begin{algorithmic}[1]
\REQUIRE edge list $E$ (undirected, $i<j$), attribute matrix $W$, number of communities $c$, maximum number of iterations $T$, relative tolerance $\tau_{\text{rel}}$
\ENSURE membership $D$, block matrix $\Theta$, attribute matrix $\Phi$, topological factor $\rho$, attribute factor $\eta$, global scale $\omega$, log-likelihood $\mathcal{L}$
\STATE Initialize $D^{(0)},\Theta^{(0)},\Phi^{(0)},\rho^{(0)},\eta^{(0)},\omega^{(0)}$ (see the initialization strategy below)
\FOR{$t = 1$ \textbf{to} $T$}
  \STATE \COMMENT{E-step}
  \STATE $M \leftarrow D\Theta,\quad P \leftarrow D\Phi$
  \FOR{each edge $(i,j) \in E$}
    \STATE $B_{ij} \leftarrow \sum_{r} M_{ir} d_{jr}$
  \ENDFOR
  \STATE Accumulate $F^{\text{raw}}$ and the topological evidence $\alpha^{\text{(link)}}$ via Eqs.~\eqref{eq:Fraw} and \eqref{eq:alpha}, grouping by node
  \STATE $\alpha^{\text{(attr)}} \leftarrow (W \oslash P)\,\Phi^{\!\top}$
  \STATE $\alpha \leftarrow \alpha^{\text{(link)}} + \alpha^{\text{(attr)}}$
  \STATE \COMMENT{M-step}
  \STATE $U \leftarrow D^{\!\top}\rho$
  \STATE $\rho \leftarrow \mathbf{g} \oslash (\omega \cdot D(\Theta U))$, then clip $\rho \leftarrow \max(\rho, \varepsilon)$
  \STATE Update $\eta$ by solving Eq.~\eqref{eq:eta} (bisection as in Algorithm~\ref{alg:bisect})
  \STATE $\Theta \leftarrow \Theta \odot F^{\text{raw}}$; $\Theta \leftarrow \Theta / \sum_{p,q}\Theta_{pq}$
  \STATE $\Phi \leftarrow \Phi \odot (D^{\!\top}(W \oslash P))$; $\Phi \leftarrow \Phi \oslash \text{rowSums}(\Phi)$
  \STATE $U \leftarrow D^{\!\top}\rho$; $C \leftarrow \Theta U$
  \FOR{$r = 1$ \textbf{to} $c$}
    \STATE Find $\lambda_r$ by bisection (Algorithm~\ref{alg:bisect}) such that $\sum_i \frac{\rho_i d_{ir}\alpha_{ir}}{\rho_i(\omega C_r + \lambda_r) + \eta_i} = 1$
    \STATE $\mathbf{d}_{*r} \leftarrow \mathbf{d}_{*r} \odot \boldsymbol{\alpha}_{*r} \oslash (\rho(\omega C_r + \lambda_r) + \eta)$
  \ENDFOR
  \FOR{$r = 1$ \textbf{to} $c$}
    \STATE $\mathbf{d}_{*r} \leftarrow \mathbf{d}_{*r} / (\rho^{\!\top}\mathbf{d}_{*r})$ \COMMENT{project onto the constraint}
  \ENDFOR
  \STATE Update $\omega$ using the new $\rho, D, \Theta$ (Eq.~\eqref{eq:omega})
  \STATE Compute $\mathcal{L}^{(t)}$ by Eq.~\eqref{eq:loglik}
  \IF{$|\mathcal{L}^{(t)} - \mathcal{L}^{(t-1)}| < \tau_{\text{rel}} (|\mathcal{L}^{(t-1)}| + 1)$ \textbf{or} $t = T$}
    \STATE \textbf{break}
  \ENDIF
\ENDFOR
\STATE \RETURN $D,\Theta,\Phi,\rho,\eta,\omega,\mathcal{L}$
\end{algorithmic}
\end{algorithm}

\begin{algorithm}[!htbp]
\caption{Bisection method for the Lagrange multipliers ($\lambda_r$ or $\nu$)}\label{alg:bisect}
\begin{algorithmic}[1]
\REQUIRE parameters defining the strictly decreasing function $f(\mu)$ on $(0,\infty)$ with $f(0^+) = +\infty$; tolerance $\varepsilon$
\ENSURE root $\mu^* > 0$
\STATE Apply a variable transformation to shift the pole to zero (Appendix~\ref{app:bisect})
\STATE Set $\mu_{\text{lo}} \leftarrow \varepsilon$, $\mu_{\text{hi}} \leftarrow 1.0$
\WHILE{$f(\mu_{\text{hi}}) \ge 0$}
  \STATE $\mu_{\text{hi}} \leftarrow 2\mu_{\text{hi}}$
\ENDWHILE
\WHILE{$f(\mu_{\text{lo}}) \le 0$}
  \STATE $\mu_{\text{lo}} \leftarrow (\mu_{\text{lo}} + \mu_{\text{hi}})/2$
\ENDWHILE
\REPEAT
  \STATE $\mu_{\text{mid}} \leftarrow (\mu_{\text{lo}} + \mu_{\text{hi}})/2$
  \IF{$f(\mu_{\text{mid}}) > 0$}
    \STATE $\mu_{\text{lo}} \leftarrow \mu_{\text{mid}}$
  \ELSE
    \STATE $\mu_{\text{hi}} \leftarrow \mu_{\text{mid}}$
  \ENDIF
\UNTIL{$|f(\mu_{\text{mid}})| < \varepsilon$}
\STATE \RETURN $\mu_{\text{mid}}$
\end{algorithmic}
\end{algorithm}

For $\lambda_r$, the function $f(\mu)$ is $g(\mu)$ in Eq.~\eqref{eq:B2}; for
$\nu$, it is $h(t)$ in Eq.~\eqref{eq:B4}. If the root does not exist (extremely
weak evidence), a large positive value is returned in order to shrink the
corresponding column of $D$.

\textbf{Initialization strategy.}
$D$ is initialized randomly and projected onto the constraints; $\Theta$ and
$\Phi$ are generated randomly and normalized; $\rho$ is initialized as the
normalized degree plus one; $\eta$ is initialized as the total attribute count
divided by its mean; and $\omega=2m$. Multiple random starts are used to avoid
local optima.

\section{Experiments and evaluation}

The experimental part of this section proceeds on two levels, synthetic networks
and real-world networks, so as to comprehensively evaluate the behavior of the
model. The methods used for comparison mainly cover three classes of
probabilistic generative models:

\begin{itemize}
\item No degree correction: PSB\_PG \cite{Chang2019};
\item Topological degree correction only: PPSB\_DC \cite{Chai2013},
NEMBP \cite{He2017}, CRSBM \cite{Ren2023};
\item Topological and attribute shared degree correction: DPSB\_PG
\cite{Zheng2020}, BCSBM \cite{Wang2025}.
\end{itemize}

The evaluation metric is the max-normalized mutual information (NMI),
\begin{equation*}
\text{NMI}_{\max}(U, V) = \frac{I(U; V)}{\max(H(U), H(V))},
\end{equation*}
where $I(U; V)$ is the mutual information between the ground-truth labels $U$ and
the clustering result $V$, and $H(\cdot)$ denotes the Shannon entropy.
$\text{NMI}_{\max}(U, V) \in [0,1]$, and the closer the value is to $1$, the
higher the agreement between the two partitions.

For the new method \DCtwo, the convergence threshold is set to $10^{-6}$ and the
maximum number of iterations to $2000$. Each network is run 10 times, and the mean
and the maximum of the 10 NMI values are reported. To guarantee a fair
comparison, all competing methods are implemented according to the algorithms in
their original papers or by using the authors' original code, with parameter
settings following the recommended values of the original papers, and all
methods are evaluated on the same data and with the same evaluation metric.

\subsection{Experiments on synthetic networks}

Through synthetic networks, controlling the network structure type, the degree of
node heterogeneity and the attribute quality, we examine the structure-detection
capability of \DCtwo\ and the effectiveness of the dual correction. We first
conduct experiments on attributed networks containing traditional community
structure, then on networks with non-traditional community structure, and finally
compare with the methods described above.

\subsubsection{Experiments on networks with traditional community structure: the case of LFR networks}

To systematically evaluate the behavior of \DCtwo\ under twofold node
heterogeneity, we use the LFR benchmark to generate topological structures with a
controllable mixing coefficient and independently design the attribute activity
factor $\etatrue$ to control its correlation with the topological degree.
Attributes are generated from community-specific Poisson distributions scaled by
$\etatrue$, as detailed below.

\textbf{Topological information generation.} The parameters for generating the
LFR network are
\[
\texttt{-N}~500 \quad \texttt{-k}~20 \quad \texttt{-maxk}~50 \quad \texttt{-mu}~0.1\sim0.9 \quad \texttt{-minc}~50 \quad \texttt{-maxc}~100,
\]
with the other parameters left at their defaults; in particular, the default
power-law exponent of the node degree distribution is 2. In this way we obtain a
topological network with node heterogeneity. To obtain attribute information with
heterogeneity, we design it as follows.

\textbf{Attribute generation model.} Define $c$ communities. We set $K = 100$,
and the true attribute distribution $\phi_r$ of each community (a $K$-dimensional
probability vector) is generated from a Dirichlet prior $\mathrm{Dir}(\beta/K)$
with $\beta = 30$, so that $\beta/K = 30/100 = 0.3 < 1$, which concentrates the
distribution on a few attributes and thus approximates the sparsity of real
labels/word frequencies. The expected attribute count of node $i$ is
\[
\hat W_{ik} = \etatrue \sum_r d_{ir}^{\text{true}} \phi_{rk},
\]
where $d_{ir}^{\text{true}}$ is the hard label (taken from the true LFR
partition) and $\etatrue$ is the attribute activity of node $i$, i.e., the
attribute heterogeneity factor we design. Finally, $W_{ik}$ is generated from a
Poisson distribution, and the total count is proportional to $\etatrue$, which
guarantees attribute abundance heterogeneity.

\textbf{Design of the attribute abundance factor $\etatrue$.} To verify the
decoupling effect of the dual correction, we control the statistical
correlation between $\etatrue$ and the topological degree $g_i$ and set up the
following three modes, listed in Table~\ref{tbl:etaModes}.

\begin{table}[htbp]
\caption{Statistical correlation between the attribute abundance factor
$\etatrue$ and the topological degree $g_i$.}\label{tbl:etaModes}
\begin{tabular}{@{}lp{0.44\linewidth}p{0.28\linewidth}@{}}
\toprule
Scenario & Generation scheme & Remark\\
\midrule
\textbf{Positive} & $\etatrue \propto g_i$ & nodes with high degree also have abundant attributes\\
\textbf{Independent} & $\etatrue$ drawn from a power-law distribution unrelated to degree, then randomly assigned to the network after shuffling the node order & tests the necessity of the dual correction\\
\textbf{Negative} & $\etatrue \propto (g_{\max} - g_i + 1)$ & low-degree nodes nevertheless have extremely abundant attributes\\
\bottomrule
\end{tabular}
\end{table}

The concrete procedure for generating $\etatrue$ is as follows.

\begin{itemize}
\item A sequence is produced from the power-law distribution
$p(\eta) \propto \eta^{-\gamma_\eta}$ ($\gamma_\eta = 2.5$) and normalized to have
mean 1.
\item For the positive and negative scenarios, the $\eta$ sequence is assigned in
the corresponding order after sorting by degree; for the independent scenario, it
is randomly permuted before assignment.
\end{itemize}

To avoid degeneracy, the lower bound of $\etatrue$ in the negative scenario is
truncated at $0.5$. The reason is that if $g_i = \max_j g_j$, then
$\eta_i \propto (\max_j g_j - \max_j g_j + 1) = 1$ and
$\eta_i^{\text{true}} = \frac{1/(\max_j g_j+1)}{1}$ (relative to the lowest-degree
node); in this case $\etatrue$ may be far smaller than $1$, so that the expected
number of attributes of node $i$ is below $1$ and the attributes generated for
that node by the Poisson distribution may all be $0$. It then becomes impossible
to distinguish whether the node genuinely has no attribute or whether its
abundance factor is squeezed to an extremely low value, so that no attribute is
observed.

\begin{figure}[htbp]
\centering
\begin{minipage}{0.47\linewidth}\centering
\includegraphics[width=\linewidth]{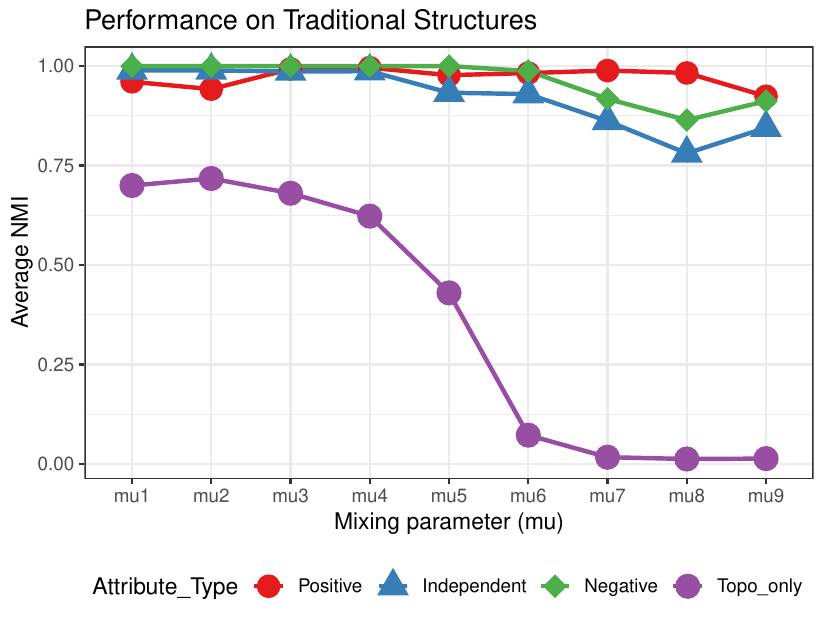}\\
{\footnotesize (a) Average NMI}
\end{minipage}\hfill
\begin{minipage}{0.47\linewidth}\centering
\includegraphics[width=\linewidth]{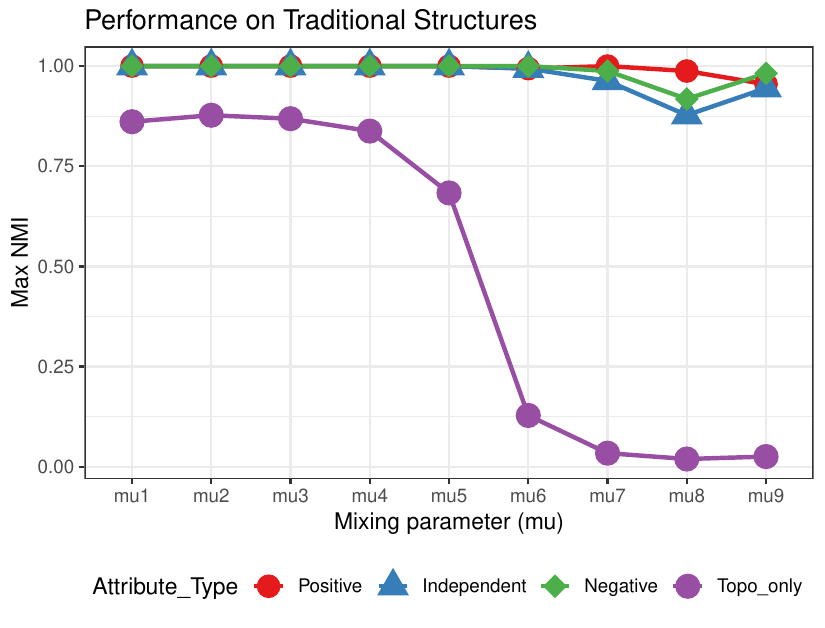}\\
{\footnotesize (b) Maximum NMI}
\end{minipage}
\caption{Community detection performance on networks with traditional community
structure; the horizontal axis is the mixing coefficient $\mu$ and the vertical
axis is the average NMI or the maximum NMI. Different colors and point shapes
denote different attribute configurations: positive correlation (Positive),
independent (Independent), negative correlation (Negative) and topology only
(Topo-only).}
\label{fig:lfr}
\end{figure}

The experimental results on the attributed networks of this part are shown in
Fig.~\ref{fig:lfr}.

It can be seen from Fig.~\ref{fig:lfr} that, in terms of both the maximum NMI and
the average NMI, the negative correlation between attribute abundance and node
degree (Negative) remains the highest under almost all values of the mixing
parameter $\mu$. In the negative-correlation scenario, topology and attributes
provide highly inconsistent but individually strong community signals. Topology
identifies communities through the edge pattern, while attributes identify
communities through the word-frequency/label pattern; both point to the same true
partition, only with opposite activity patterns. The experimental results show
that \DCtwo\ can capture both signals independently and that, because there is no
``compromise'' through a shared factor, the signals are not diluted but instead
strengthened in both channels. Therefore, even when $\mu$ is large and the
topological signal is extremely weak, the attribute signal can still
independently support accurate community discovery.

The positive-correlation configuration (Positive) is slightly better than the
independent one (Independent) when $\mu$ is large. This is because, when attribute
abundance is positively correlated with node degree, the two channels make a
consistent judgement about node ``activity'': high-degree nodes also possess more
attributes. This consistency means that the model does not need to coordinate the
two correction factors additionally, and the estimates of $\rho_i$ and $\eta_i$
are more stable; hence, even when the topological signal decays, the attribute
signal not only compensates for it but also acts synergistically with the
residual topological signal, further improving the accuracy.

In the independent case (Independent), the stability is slightly inferior to the
correlated scenarios, probably because attributes and degree are completely
independent, so that node activity is unrelated across the two channels and the
model must infer $\rho_i$ and $\eta_i$ separately from scratch; although the
decoupling of the parameter space is thorough, it also increases the difficulty
of optimization. Nevertheless, even in the worst case its average NMI still far
exceeds that of the topology-only method, which proves that the attribute
information itself has been effectively exploited.

In summary, the experimental results show that the independent dual-correction
mechanism of \DCtwo\ can fully decouple the heterogeneity of nodes along the two
dimensions of topology and attributes, avoiding the performance loss caused by
the confusion of the two kinds of activity in traditional methods. The
topology-only method fails rapidly once the mixing parameter exceeds $0.5$,
further highlighting the necessity of fusing attribute information and the
efficiency with which \DCtwo\ exploits it.

\subsubsection{Experiments on networks with generalized community structure: the case of networks generated by the DCSBM}

To evaluate the ability of \DCtwo\ to identify non-traditional, non-assortative
community structures, in particular its adaptability to disassortative,
bipartite and mixed structures, we design the following experiment.

The number of nodes is $n = 500$ and the number of communities is $c = 4$. The
node degree correction factor $\rho_i \sim \text{Gamma}(2, 2)$ is then linearly
scaled to have mean $1$, producing moderate heterogeneity (coefficient of
variation $\approx 0.707$). The block matrix $\Theta$ is designed with three
structures:

\begin{itemize}
\item Disassortative structure: connections within a community are extremely
weak (diagonal $0.1$) and connections across communities are strong
(off-diagonal $1.0$), i.e.,
$\Theta_{\text{dis}} = 0.1 \cdot I + 1.0 \cdot (\mathbf{1}\mathbf{1}^\top - I)$.
\item Bipartite structure: the communities are divided into two groups
$G_1 = \{1,2\}$ and $G_2 = \{3,4\}$, with extremely weak intra-group connections
($0.01$) and strong inter-group connections ($1.0$). Every pair of cross-group
communities is densely connected, exhibiting the typical characteristics of a
bipartite graph; that is, $\Theta_{\text{bi}}$ equals $0.01$ when $z_i,z_j$ both
belong to $\{1,2\}$ or both to $\{3,4\}$, and $1.0$ otherwise.
\item Mixed structure. To keep the same scale as the previous experiments
($c=4$), we design a mixed block matrix with asymmetric strengths, as follows:
\[
\Theta_{\text{mixed}} = \begin{pmatrix}
5 & 1 & 0.1 & 3 \\
1 & 5 & 3 & 0.1 \\
0.1 & 3 & 5 & 1 \\
3 & 0.1 & 1 & 5
\end{pmatrix}.
\]
\end{itemize}

In this matrix the diagonal elements are $5$ (relatively strong connections
within a community, similar to assortativity), while the off-diagonal elements
take a variety of values: communities 1 and 2 are moderately connected ($1$),
communities 1 and 4 are relatively strongly connected ($3$), and communities 1
and 3 are very weakly connected ($0.1$). The overall pattern is asymmetric
(although the matrix itself is symmetric, the patterns of the different
rows/columns differ), and it cannot be simply classified as assortative,
disassortative or bipartite.

For networks of all three structures, the adjacency matrix is generated as
follows: for all $i \ne j$,
$A_{ij} \sim \text{Poisson}(\rho_i \rho_j \Theta_{z_i, z_j})$, ignoring
self-loops. Multi-edges arise naturally but are very rare in the sparse setting.
The scaling factor is searched for by the algorithm during the iterations, so as
to match the total number of edges of the networks with different structures
exactly. The design of the attribute information is the same as in the previous
subsection.

\begin{figure}[htbp]
\centering
\begin{minipage}{0.47\linewidth}\centering
\includegraphics[width=\linewidth]{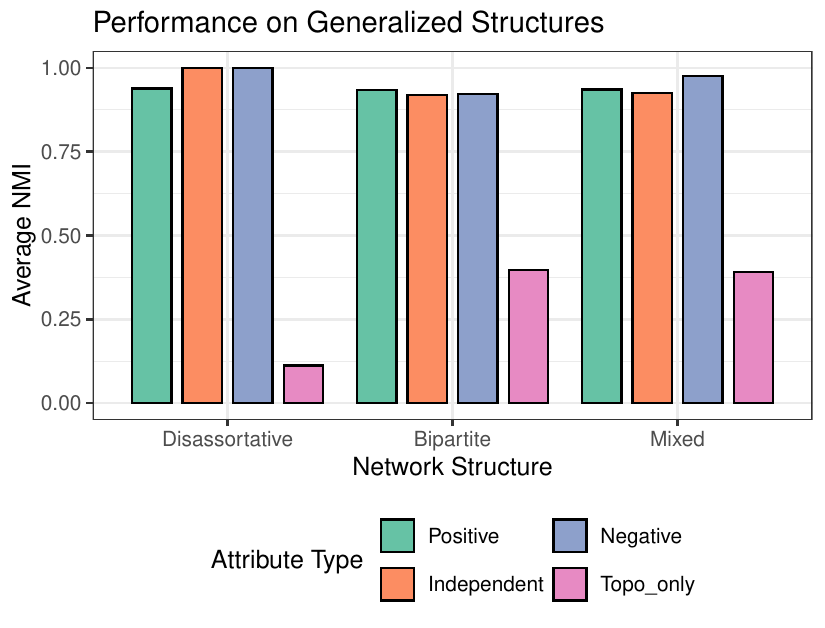}\\
{\footnotesize (a) Average NMI}
\end{minipage}\hfill
\begin{minipage}{0.47\linewidth}\centering
\includegraphics[width=\linewidth]{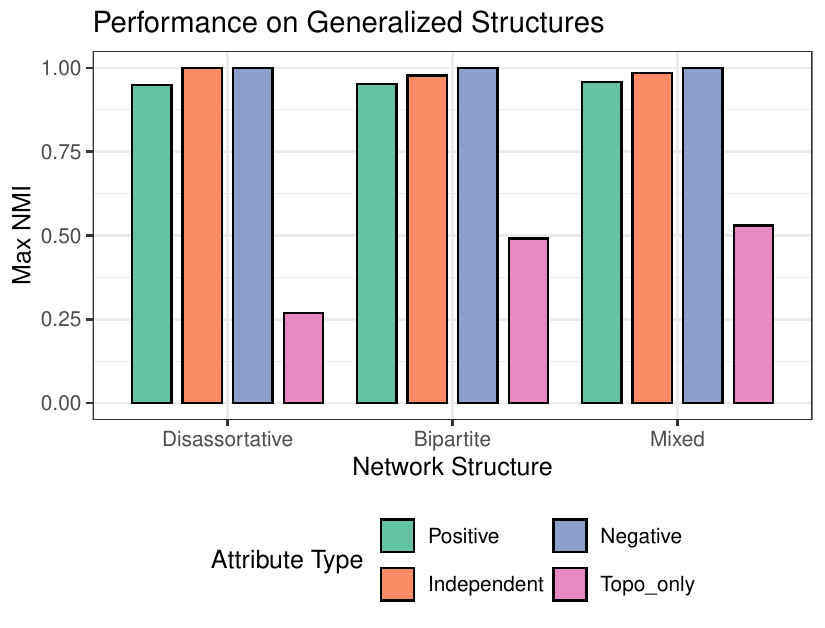}\\
{\footnotesize (b) Maximum NMI}
\end{minipage}
\caption{Community detection performance under different generalized structures;
the horizontal axis is the network structure type and the vertical axis is the
average NMI and the maximum NMI. Different fill colors denote the four attribute
configurations: positive correlation (Positive), independent (Independent),
negative correlation (Negative) and topology only (Topo-only).}
\label{fig:sbm}
\end{figure}

The experimental results on the attributed networks of this part are shown in
Fig.~\ref{fig:sbm}.

From Fig.~\ref{fig:sbm} it can be seen that the average NMI or the maximum NMI of
the topology-only method (Topo-only) is far lower than the NMI values of the
attribute-aware methods, indicating that the model can capture, from the attribute
information, community structure that the topology alone cannot reflect. On the
disassortative structure, Negative and Independent both reach a perfect
$\text{NMI} = 1.000$, and they are also at the highest level on the bipartite and
mixed structures. This visual observation directly confirms the core design of
this paper: the independent dual-correction factors $\rho_i$ and $\eta_i$ can
decouple the activity differences of a node along the two dimensions of topology
and attributes. Even when attribute abundance is independent of, or inversely
related to, topological degree, the model can still extract the correct
community information from both channels simultaneously, without being disturbed
by the conflict between the activity patterns.
\subsubsection{Comparison of the methods on synthetic networks}

For the synthetic networks containing traditional communities we select the
network corresponding to $\mu = 0.7$ in Section~5.1.1, whose topological signal is
weak; for the non-traditional networks we select the three structures of
Section~5.1.2. Each network corresponds to the three attribute configurations
(positive, independent and negative), giving 12 synthetic networks in total. The
average NMI over 10 runs of each method on each network and its standard
deviation are reported in Table~\ref{tbl:synthetic}. Because PPSB\_DC
\cite{Chai2013} was run 30 times on each network in the original paper, we follow
the same procedure here and report the mean and standard deviation of the 10 runs
with the largest likelihood.

\begin{table}[htbp]
\caption{Comparison on the synthetic networks (mean$\pm$sd). A7 denotes an LFR
network with $\mu=0.7$ (Section~5.1.1); Adis, Abi and Amix denote the
disassortative, bipartite and mixed structures of Section~5.1.2; ``posi'', ``indep''
and ``neg'' correspond to the positive, independent and negative attribute
configurations, respectively. The best results are in bold.}\label{tbl:synthetic}
{\footnotesize\setlength{\tabcolsep}{3pt}%
\begin{tabular}{@{}lccccccc@{}}
\toprule
Networks & PPSB\_DC & NEMBP & PSB\_PG & DPSB\_PG & CRSBM & BCSBM & \DCtwo\\
\midrule
A7posi      & 0.5688$\pm$0.1176 & 0.8633$\pm$0.0111 & 0.9691$\pm$0.0589 & 0.9877$\pm$0.0000 & 0.0462$\pm$0.0293 & 0.9877$\pm$0.0000 & \textbf{0.9898$\pm$0.0056}\\
A7indep     & 0.4754$\pm$0.0612 & 0.8286$\pm$0.0905 & 0.5900$\pm$0.1643 & 0.8106$\pm$0.0565 & 0.0220$\pm$0.0090 & 0.8104$\pm$0.0565 & \textbf{0.8708$\pm$0.0628}\\
A7neg       & 0.6825$\pm$0.1118 & 0.8458$\pm$0.0924 & 0.6215$\pm$0.1843 & 0.9058$\pm$0.0139 & 0.1143$\pm$0.0275 & 0.9058$\pm$0.0139 & \textbf{0.9146$\pm$0.0603}\\
Adis\_posi  & 0.5862$\pm$0.0211 & 0.7608$\pm$0.1771 & \textbf{0.9505$\pm$0.0034} & \textbf{0.9505$\pm$0.0041} & 0.0125$\pm$0.0119 & \textbf{0.9505$\pm$0.0041} & 0.9389$\pm$0.0067\\
Adis\_indep  & 0.5318$\pm$0.0179 & 0.7552$\pm$0.1508 & 0.7074$\pm$0.1558 & \textbf{1.0000$\pm$0.0000} & 0.0185$\pm$0.0202 & \textbf{1.0000$\pm$0.0000} & \textbf{1.0000$\pm$0.0000}\\
Adis\_neg    & 0.5562$\pm$0.0246 & 0.8499$\pm$0.1292 & 0.8299$\pm$0.1797 & 0.9992$\pm$0.0027 & 0.1026$\pm$0.0450 & 0.9992$\pm$0.0027 & \textbf{1.0000$\pm$0.0000}\\
Abi\_posi   & 0.5033$\pm$0.0226 & 0.8607$\pm$0.1087 & 0.9591$\pm$0.0058 & \textbf{0.9596$\pm$0.0055} & 0.0381$\pm$0.0160 & \textbf{0.9596$\pm$0.0055} & 0.9334$\pm$0.0109\\
Abi\_indep   & 0.6595$\pm$0.0264 & 0.9000$\pm$0.1291 & 0.6311$\pm$0.0804 & 0.8287$\pm$0.0143 & 0.0245$\pm$0.0161 & 0.8281$\pm$0.0152 & \textbf{0.9189$\pm$0.0386}\\
Abi\_neg     & 0.7271$\pm$0.0266 & 0.8500$\pm$0.1291 & 0.9096$\pm$0.1407 & 0.9914$\pm$0.0093 & 0.1030$\pm$0.0323 & \textbf{0.9936$\pm$0.0089} & 0.9222$\pm$0.0486\\
Amix\_posi  & 0.8990$\pm$0.0304 & 0.7941$\pm$0.1189 & \textbf{0.9588$\pm$0.0026} & 0.9580$\pm$0.0000 & 0.4203$\pm$0.0425 & 0.9580$\pm$0.0000 & 0.9358$\pm$0.0169\\
Amix\_indep  & 0.8075$\pm$0.0094 & \textbf{0.9750$\pm$0.0791} & 0.6121$\pm$0.0161 & 0.8599$\pm$0.0014 & 0.3709$\pm$0.0530 & 0.8597$\pm$0.0017 & 0.9251$\pm$0.0989\\
Amix\_neg    & 0.7164$\pm$0.0140 & 0.9242$\pm$0.1202 & 0.8372$\pm$0.1721 & \textbf{1.0000$\pm$0.0000} & 0.4122$\pm$0.0431 & \textbf{1.0000$\pm$0.0000} & 0.9752$\pm$0.0582\\
\bottomrule
\end{tabular}}
\end{table}

It can be seen from Table~\ref{tbl:synthetic} that on the 12 networks the new
method takes the lead on the largest number of networks, namely 6. It is followed
by BCSBM and DPSB\_PG, which lead on 5 and 4 networks, respectively. The former
uses the composite node influence as the correction factor and the latter uses
the node degree; the two methods differ mainly in the input factor. A further
inspection reveals that the two methods give completely identical results on many
networks, which may be related to the fact that their algorithms differ only in the
input factor.

\subsection{Experiments on real-world networks}

Synthetic data cannot fully reflect the complexity of real network structures and
the heterogeneity of node degree and attribute abundance. To test the behavior of
the proposed method more comprehensively, we select real attributed networks with
different structures and sizes. The real attributed networks used include WebKB
(Cornell, Texas, Washington, Wisconsin), Cora, CiteSeer and LastFM (LastFMsmall,
LastFMmedium, LastFMlarge, LastFMhuge).

\begin{itemize}
\item WebKB dataset: this dataset contains four sub-networks, Cornell, Texas,
Washington and Wisconsin. Each sub-network is divided into 5 communities; the four
sub-networks collectively contain 877 web pages and 1608 edges, and each web page
is described by 1703 binary attributes.
\item LastFM dataset: a social network of LastFM users whose nodes are users
and whose edges are mutual follower relationships; the node features are
extracted from the artists liked by the users. In \cite{Hu2024}, the original
data were organized into four subsets according to the number of users per
country: small ($<100$ users), medium (100--300), large (300--1000) and huge
($>1000$), corresponding to LastFMsmall, LastFMmedium, LastFMlarge and
LastFMhuge, respectively. Since isolated nodes are present in these subsets,
we remove them before use; see Table~\ref{tbl:datasets} for details.
\item Cora dataset: the Cora dataset consists of 2708 scientific publications
classified into one of seven classes. The citation network consists of 5429
links. Each publication is described by a 0/1-valued word vector indicating the
absence/presence of the corresponding word from the dictionary; the dictionary
consists of 1433 unique words.
\item CiteSeer dataset: the CiteSeer dataset consists of 3312 scientific
publications classified into one of six classes. The citation network consists of
4732 links. Each publication is described by a 0/1-valued word vector indicating
the absence/presence of the corresponding word from the dictionary; the
dictionary consists of 3703 unique words.
\end{itemize}

A detailed description of the networks is given in Table~\ref{tbl:datasets}.

\begin{table}[htbp]
\caption{Basic description of the real-world networks. ``Networks'' is the
network name, $n$ is the number of nodes, $m$ is the number of edges, $K$ is the
attribute dimension, $c$ is the number of communities, and ``Structure'' is the
network structure type (mixture = mixed structure, community = community
structure; a missing marker is denoted by ``---'').}\label{tbl:datasets}
{\footnotesize\setlength{\tabcolsep}{5pt}%
\begin{tabular}{@{}lccccc@{}}
\toprule
Networks & $n$ & $m$ & $K$ & $c$ & Structure\\
\midrule
cornell      & 195  & 304   & 1703 & 5 & mixture\\
texas        & 187  & 328   & 1703 & 5 & mixture\\
washington   & 230  & 446   & 1703 & 5 & mixture\\
wisconsin    & 265  & 530   & 1703 & 5 & mixture\\
LastFMsmall  & 308  & 832   & 194  & 6 & ---\\
LastFMmedium & 539  & 2212  & 324  & 3 & ---\\
LastFMlarge  & 2330 & 7738  & 600  & 5 & ---\\
LastFMhuge   & 3570 & 13052 & 600  & 3 & ---\\
cora         & 2708 & 5429  & 1433 & 7 & community\\
citeseer     & 3312 & 4723  & 3703 & 6 & community\\
\bottomrule
\end{tabular}}
\end{table}

The average NMI over 10 runs of the new model on the real-world networks and its
standard deviation are reported in Table~\ref{tbl:realresults}, together with the
average metrics of the competing methods.

\begin{table}[htbp]
\caption{Results on the real-world networks (mean$\pm$sd).Best results are in bold.}\label{tbl:realresults}
{\footnotesize\setlength{\tabcolsep}{3pt}%
\begin{tabular}{@{}lccccccc@{}}
\toprule
Networks & PPSB\_DC & NEMBP & PSB\_PG & DPSB\_PG & CRSBM & BCSBM & \DCtwo\\
\midrule
cornell      & 0.1334$\pm$0.0151 & 0.1647$\pm$0.0501 & 0.3137$\pm$0.0567 & 0.3311$\pm$0.0505 & 0.1006$\pm$0.0230 & 0.3263$\pm$0.0504 & \textbf{0.3438$\pm$0.0509}\\
texas        & 0.2370$\pm$0.0524 & 0.2536$\pm$0.0383 & 0.2358$\pm$0.0546 & 0.2657$\pm$0.0488 & 0.0987$\pm$0.0347 & 0.2668$\pm$0.0316 & \textbf{0.2861$\pm$0.0358}\\
washington   & 0.1634$\pm$0.0260 & 0.1456$\pm$0.0682 & 0.2743$\pm$0.0330 & 0.2960$\pm$0.0256 & 0.1369$\pm$0.0429 & 0.2960$\pm$0.0256 & \textbf{0.3210$\pm$0.0504}\\
wisconsin    & 0.1146$\pm$0.0095 & 0.0727$\pm$0.0189 & 0.3458$\pm$0.0489 & 0.3404$\pm$0.0292 & 0.0814$\pm$0.0315 & 0.3404$\pm$0.0292 & \textbf{0.3544$\pm$0.0334}\\
LastFMsmall  & 0.5040$\pm$0.0237 & \textbf{0.7849$\pm$0.0389} & 0.4179$\pm$0.0642 & 0.4208$\pm$0.0909 & 0.6905$\pm$0.0589 & 0.4126$\pm$0.0866 & 0.4262$\pm$0.0379\\
LastFMmedium & 0.7635$\pm$0.1848 & 0.7984$\pm$0.1549 & 0.8352$\pm$0.0123 & 0.8389$\pm$0.0056 & 0.6374$\pm$0.1422 & 0.8395$\pm$0.0037 & \textbf{0.8418$\pm$0.0059}\\
LastFMlarge  & \textbf{0.6027$\pm$0.0119} & 0.5520$\pm$0.0243 & 0.4961$\pm$0.0400 & 0.4786$\pm$0.0225 & 0.3753$\pm$0.0694 & 0.4743$\pm$0.0244 & 0.4654$\pm$0.0444\\
LastFMhuge   & 0.8571$\pm$0.0000 & \textbf{0.9301$\pm$0.0130} & 0.9016$\pm$0.0047 & 0.8905$\pm$0.0024 & 0.3320$\pm$0.1513 & 0.8911$\pm$0.0025 & 0.8946$\pm$0.0046\\
cora         & 0.2214$\pm$0.0059 & \textbf{0.3851$\pm$0.0521} & 0.2382$\pm$0.0540 & 0.3264$\pm$0.0286 & 0.3058$\pm$0.0389 & 0.3212$\pm$0.0256 & 0.2952$\pm$0.0430\\
citeseer     & \textbf{0.3506$\pm$0.0330} & 0.1980$\pm$0.0441 & 0.2906$\pm$0.0397 & 0.2922$\pm$0.0399 & 0.1674$\pm$0.0296 & 0.2902$\pm$0.0392 & 0.2754$\pm$0.0416\\
\bottomrule
\end{tabular}}
\end{table}

It can be seen from Table~\ref{tbl:realresults} that the new method achieves the
highest average NMI on 5 networks, including the WebKB datasets and LastFMmedium,
and is inferior to NEMBP and PSB\_PG on LastFMhuge, although it still outperforms
the remaining methods. If the two correction factors of \DCtwo\ are merged into a
single factor fixed to the node degree, the model reduces to a method similar to
DPSB\_PG or BCSBM; comparing only these three models, the new model shows a
certain advantage on 7 networks, including the WebKB datasets, LastFMsmall,
LastFMmedium and LastFMhuge. Unlike its behavior on the synthetic networks, NEMBP
also performs prominently on the real networks, achieving the highest average NMI
on LastFMsmall, LastFMhuge and cora, which indicates that the structure of real
networks still differs to some extent from that of synthetic networks. Overall,
however, on both synthetic and real networks the new method demonstrates strong
competitiveness compared with existing state-of-the-art methods.

\subsection{Recoverability, decoupling and necessity of the correction factors}
\label{sec:factors}

Sections 5.1 and 5.2 evaluate community detection quality via NMI, which measures how well the estimated membership matrix $D$ reproduces the ground-truth partition. NMI, however, only quantifies agreement between two discrete partitions; it does not assess whether the continuous node-heterogeneity factors $\rho_i$ and $\eta_i$ are correctly recovered. The appropriate criterion is therefore the similarity between the inferred factors and their ground truths. Although MSE or RMSE might seem natural, they are sensitive to the global scale ambiguity of $\rho$ and $\eta$ and to the heavy-tailed Pareto distribution of $\eta$; Spearman correlation, by contrast, is scale-invariant, monotone-invariant, and robust to heavy tails. We therefore adopt it as the unified criterion throughout Section 5.3.

In the LFR networks of Section~5.1.1, the degree
heterogeneity of the nodes comes from a prescribed power-law degree sequence
rather than from an explicit multiplicative factor, so there is no ground truth
for $\rho$. The real-world networks (Section~5.2) have no ground-truth factors.
Therefore, this section uses the synthetic networks of Section~5.1.2---whose
generative process provides the ground truths $\rhotrue$ and $\etatrue$ of both
factors simultaneously---to examine the recoverability, decoupling and necessity
of the factors.

\subsubsection{Recoverability of the correction factors}
\label{sec:recover}

The generation parameters of the network and attributes are the same as in
Section~5.1.2 and are not repeated. The ground truth of the attribute abundance
factor follows the three configurations of Section~5.1.1, denoted here as
positive (Positive, $\eta_i^{\text{true}}\propto g_i$), independent (Independent,
Pareto sampling unrelated to $g$, $\gamma_\eta=2.5$) and negative (Negative,
$\eta_i^{\text{true}}\propto(\max_j g_j-g_i+1)$, lower bound 0.5). The three
configurations are applied, one by one, to networks with three different
community structures, giving 9 network instances.

Each instance is fitted with 10 different random initializations; the solution
with the largest log-likelihood is taken as the main result and the remaining
solutions are used to assess initialization stability. All 90 fits converge
under the setting of a maximum of 2000 iterations and a relative convergence
tolerance of $10^{-6}$. Because the scale conventions of the generator and of the
model differ, the ground-truth and inferred factors are each normalized to mean
1 before comparison, so that the regression slope can be compared directly with
1. The agreement between the ground-truth and inferred values is measured by the
Spearman correlation. The
results for the mixed structure are shown in Fig.~\ref{fig:recovery}; those for
the other two structures are given in Appendix~\ref{app:recovery}.

\begin{figure}[htbp]
\centering
\includegraphics[width=0.92\linewidth]{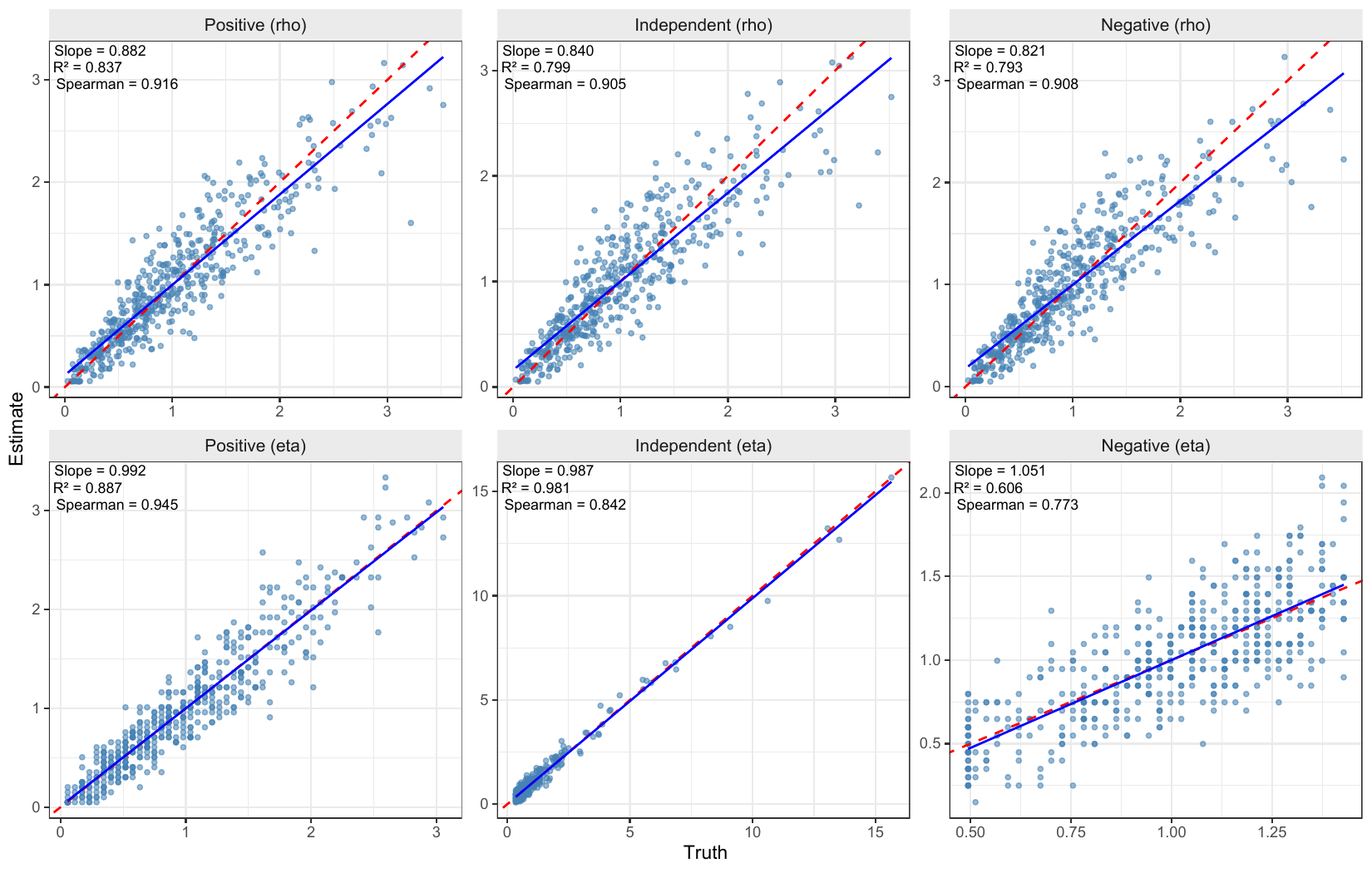}
\caption{Scatter plots of the ground-truth versus the inferred correction factors
under the mixed structure. The horizontal axis is the ground truth of the
correction factor and the vertical axis is the inferred value of the
corresponding correction factor; the upper panels are the topological degree
factor $\rho$ and the lower panels are the attribute abundance factor $\eta$; the
three columns correspond, in order, to positive (Positive), independent
(Independent) and negative (Negative). The red dashed line is $y=x$ and the blue
solid line is the least-squares fit. In the third column the ground truth of
$\eta$ is compressed to the range $[0.49,1.43]$ by the 0.5 lower-bound
truncation, so its horizontal-axis range is markedly smaller than that of the
first two columns (for the same structure, $[0.058,3.05]$ for the positive
configuration and $[0.350,15.6]$ for the independent configuration).}
\label{fig:recovery}
\end{figure}

As can be seen from Fig.~\ref{fig:recovery}, in all panels the scatter points lie
along $y=x$ (the red dashed line): taking the ground truth of the correction
factor as the independent variable and the inferred value as the dependent
variable, the least-squares slope lies between $0.82$ and $1.05$, and the
Spearman correlation lies between $0.773$ and $0.946$, indicating that the
algorithm recovers the node heterogeneity factors well. It should be noted that
in the independent configuration the ground truth of $\eta$ follows a
heavy-tailed distribution and the regression variance is dominated by a few tail
observations, so that the $R^2$ of that panel is as high as $0.981$ while the
Spearman correlation is only $0.842$---this is precisely why this paper adopts
the rank correlation as the primary measure. The Spearman correlation of $\eta$
in the negative configuration ($0.637\sim0.773$) is the lowest of the three. This
is because in that configuration the ground truth is compressed, after the 0.5
lower-bound truncation, to the range $[0.494,1.43]$, and about $4.0\%\sim6.8\%$
of the high-degree nodes have ground truths tied at the lower bound, so that the
rank information is reduced accordingly and the Spearman correlation is lowered.

As regards initialization stability, over 10 random initializations the standard
deviation of the Spearman correlation of the topological degree factor is at most
$0.161$ (mixed structure, negative configuration), and for all other instances it
does not exceed $0.064$. Even in the most sensitive mixed structure,
$\mathrm{Spearman}(\rhohat,\rhotrue)$ still reaches $0.833$ when averaged over 10
runs, so the initialization sensitivity does not change the conclusion on
recoverability. The standard deviation of the attribute abundance factor is
almost always zero.

\subsubsection{Decoupling of the correction factors}
\label{sec:decouple}

In the design, the two correction factors in the new model are determined by the
observations of their respective channels and are decoupled. If this design is
successful, the heterogeneity of one channel should not affect the factor
estimate of the other channel. This can be verified as follows: the accuracy with
which $\rhohat$ recovers $\rhotrue$ should remain unchanged as the relationship
between $\etatrue$ and the node degree $g$ is reversed from positive correlation,
through independence, to negative correlation.

\begin{table}[htbp]
\caption{Values of $\mathrm{Spearman}(\rhohat,\rhotrue)$ under different
attribute configurations on the 9 networks of Section~\ref{sec:recover}. In columns 2 to 4, positive, independent and negative
refer to the correlation configuration between $\etatrue$ and $g$.}\label{tbl:rhoDecouple}
{\footnotesize\setlength{\tabcolsep}{8pt}%
\begin{tabular}{@{}lcccc@{}}
\toprule
Structure & Positive & Independent & Negative & Within-group range\\
\midrule
Disassortative & 0.9247 & 0.9239 & 0.9249 & 0.001\\
Bipartite      & 0.9299 & 0.9300 & 0.9286 & 0.001\\
Mixed          & 0.9160 & 0.9050 & 0.9076 & 0.011\\
\bottomrule
\end{tabular}}
\end{table}

As can be seen from Table~\ref{tbl:rhoDecouple}, the relationship between the
ground truth of the attribute abundance factor and the node degree $g$ goes from
perfectly positively correlated, through independent, to perfectly negatively
correlated; yet within the same structure the range of the Spearman
rank correlation between $\rhohat$ and $\rhotrue$ does not exceed $0.011$, and
the values of all 9 instances fall in the narrow interval $0.905\sim0.930$. In
other words, although the relationship between the attribute side and the degree
is completely reversed, the recovery accuracy of the topological factor is almost
unchanged. It means that the two factors are decoupled:
the heterogeneity of the attribute side is not absorbed into the topological
correction factor. This corroborates the earlier observation in Section~5.1 that
the clustering performance does not drift with the attribute configuration
(Fig.~\ref{fig:sbm}).

\subsubsection{Necessity of the dual correction}
\label{sec:necessity}

Sections~\ref{sec:recover} and~\ref{sec:decouple} have shown, respectively, that the two factors can be inferred separately and that they are mutually decoupled. A stronger question then arises: can the two factors be merged into a single shared factor? This section constructs a controlled mismatch experiment that answers under what conditions shared modeling fails, and when this cost propagates to the clustering result.

Depending on whether the factors are \emph{adaptively updated}, shared models can be further divided into two categories: ``fixed sharing'' and ``adaptive sharing''. DPSB\_PG and BCSBM belong to the former, as they fix a single node factor at initialization and use it unchanged in both channels. The present model, by contrast, introduces two mutually independent factors, both updated adaptively by EM. Therefore, ``the present model reduces to DPSB\_PG/BCSBM'' holds strictly only when the factor is ``shared and fixed''; if the two factors are merely made to share while adaptive updating is retained, the result is another shared variant not yet found in the literature.

\textbf{Analytical upper bound of shared modeling.} We first state a fact that
does not depend on experiments. DPSB\_PG and BCSBM fix the node factor to the
observed degree $g_i$ (or to a composite indicator formed from the degree,
betweenness centrality, etc.), and the expectation on the attribute side is
$\hat W_{ik}=g_i\sum_r d_{ir}\phi_{rk}$; that is, the attribute abundance factor
is by definition proportional to $g_i$. Hence, no matter how far the iterative
optimization proceeds, we always have
\begin{equation}
\mathrm{Spearman}\!\left(\hat\eta_{\text{shared}},\etatrue\right)\equiv\mathrm{Spearman}\!\left(g,\etatrue\right).
\label{eq:shared}
\end{equation}
For the adaptive sharing variant, Eq.~\eqref{eq:shared} still holds, with $g$
replaced by the adaptively inferred topological factor $\rhohat$ of the
topological channel: its recovery accuracy is identically
$\mathrm{Spearman}(\rhohat,\etatrue)$ and likewise does not depend on the
iterative process of the attribute channel.

Furthermore, the observed degree is not the true topological factor itself but
only a noisy observation of it: the generative mechanism gives
$\mathbb E[g_i]\propto\rho_i^{\text{true}}$, and finite-sample fluctuations place
$\mathrm{Spearman}(g,\rhotrue)$ between $0.92$ and $0.94$ on the synthetic
networks of this paper (this holds for all three structures); the rank
correlation between the adaptively inferred $\rhohat$ and $\rhotrue$ is likewise
only $0.905\sim0.930$ (Section~\ref{sec:recover},
Table~\ref{tbl:rhoDecouple}). Therefore, when the true rank correlation of the
two factors is $s$, the recovery accuracy of both shared variants is
approximately $0.93\,s$---it is capped linearly by the coupling strength of the
two factors: even at full synchronization ($s=+1$) it can only reach about $0.9$;
once they decouple it declines proportionally; and at full reversal ($s=-1$) it
takes about $-0.9$. As will be seen below, the shape of the shared curves in
Fig.~\ref{fig:mismatch} is a direct manifestation of this analytical prediction,
and the near-coincidence of the two shared curves is precisely because $g$ and
$\rhohat$ have almost the same fidelity to $\rhotrue$.

\textbf{Design of the controlled experiment.} The experiment is conducted on the
synthetic networks with the three generalized community structures of
Section~5.1.2, and the generation parameters of topology and attributes are the
same as in that section and are not repeated. The only change in this section is
that the value of the rank correlation between $\etatrue$ and $\rhotrue$ is
refined from the three configurations of Section~5.1.1 (positive, independent,
negative) into an 11-level controlled sweep: the target value $s$ goes from $+1$
(the two factors fully synchronized) to $-1$ (fully reversed) with a step of
$0.2$. To make the coupling strength the only variable, we fix the network
topology (the same random seed), the community--attribute distribution $\Phi$,
the candidate value set of $\etatrue$, and all random numbers; between the 11
levels only the correspondence between candidate values and nodes is changed. The
candidate values are a fixed set of Pareto quantile grid points ($\gamma_\eta=2.5$,
normalized to mean $1$). The assignment order is generated by a Gaussian copula:
the ranks of $\rhotrue$ are transformed into standard normal quantiles $z_1$, a
latent variable $q=\rho_G z_1+\sqrt{1-\rho_G^2}\,\varepsilon$ is constructed by
adding Gaussian noise ($\varepsilon$ being standard normal noise), and the
candidate values are then assigned to the nodes in ascending order of the ranks
of $q$; the latent correlation coefficient is obtained by inverting the
rank-correlation formula of the Gaussian copula,
$s=(6/\pi)\arcsin(\rho_G/2)$, i.e., $\rho_G=2\sin(\pi s/6)$, so that the rank
correlation between $q$ and $\rhotrue$ equals the target value $s$. Hence the
value set and the marginal distribution of $\etatrue$ are exactly the same across
the 11 levels, the only difference being ``which node is assigned which value'';
at the two ends $s=\pm1$ we have $q=\pm z_1$, so $\etatrue$ degenerates into a
strictly monotone function of $\rhotrue$ and the actual rank correlation is
exactly $\pm1$. In addition, the copula noise $\varepsilon$ and the random
numbers used for the Poisson attribute sampling are shared across the 11 levels
(common random numbers), which further isolates fluctuations unrelated to the
coupling strength. The three community structures each have 11 levels, giving 33
instances in total; the deviation of the actually achieved rank correlation from
the target value $s$ does not exceed $0.04$. Each instance is fitted with 10
random initializations, and the solution with the largest log-likelihood is taken
as the main result. Under this design the minimum value of $\etatrue$ is about
$1/3$, and the expected attribute count of each node is still far greater than
$1$, so the degeneracy discussed in Section~5.1.1---attributes becoming all zero
because the expected count is too low---does not occur, and no lower-bound
truncation is needed.

\begin{figure}[htbp]
\centering
\includegraphics[width=0.98\linewidth]{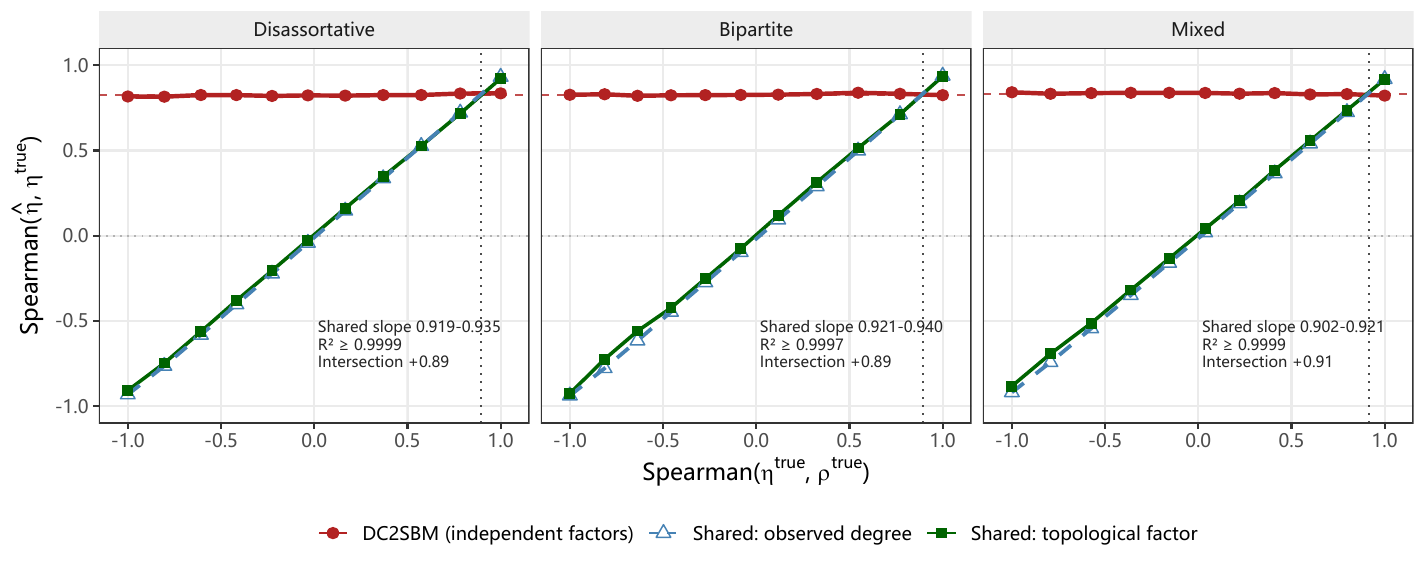}
\caption{Variation of the recovery accuracy of the attribute abundance factor
with the coupling strength of the two factors in the controlled mismatch
experiment. The horizontal axis is the actually achieved
$\mathrm{Spearman}(\etatrue,\rhotrue)$ (deviating from the design target by no
more than $0.04$) and the vertical axis is the recovery accuracy
$\mathrm{Spearman}(\etahat,\etatrue)$; the three panels correspond, in order, to
the disassortative, bipartite and mixed structures. The red circles are the
experimental results of \DCtwo\ (independent inference of the two factors); the
blue triangles are the fixed sharing variant (attribute abundance factor fixed to
the observed degree $g$, i.e., the representative practice of DPSB\_PG and
BCSBM); and the green squares are the adaptive sharing variant (attribute
abundance factor taken as the adaptively inferred $\rhohat$ of the topological
channel). The grey dashed line is the $0$ reference line. For each level, the
solution with the largest log-likelihood among 10 random initializations is
used.}\label{fig:mismatch}
\end{figure}

\textbf{Results.} As shown in Fig.~\ref{fig:mismatch}, the two classes of
modeling exhibit a consistent yet sharply different shape under the three
structures. The two shared curves vary approximately linearly with the coupling
strength, with least-squares slopes of $0.902\sim0.940$, in close agreement with
the analytical prediction $0.93\,s$; whereas the curve of \DCtwo\ remains
horizontal across all 11 levels, with a slope of at most $0.008$ in absolute
value and a cross-level excursion of only $0.018\sim0.020$, the residual
fluctuations being at the level of initialization noise. The two curves intersect
at about $+0.89\sim+0.91$ on the horizontal axis: only when the coupling strength
of the two factors exceeds this value is shared modeling no worse than the
independent two-factor scheme; once the two decouple, the recovery accuracy of
shared modeling declines linearly with the coupling strength. At the most
unfavourable end (the two factors fully reversed), the recovery accuracy of the
shared class is $-0.882\sim-0.938$---it not only fails to recover but gives a
factor whose direction is entirely wrong---whereas \DCtwo\ still gives
$+0.817\sim+0.842$. In addition, the slopes ($0.902\sim0.921$ and
$0.921\sim0.940$) and the excursions ($1.800\sim1.861$ and $1.838\sim1.876$) of
the fixed-sharing and adaptive-sharing curves are almost identical, and the two
curves essentially coincide in the figure, which indicates that the root of the
failure lies in the modeling choice of ``sharing the two factors'' itself and is
independent of which statistic is shared and of whether that statistic is fixed
or adaptively updated. Combined with the evidence in
Sections~\ref{sec:recover} and~\ref{sec:decouple} that the two factors can be
recovered separately, it can be concluded that, in networks where attribute
abundance and topological activity are not necessarily synchronized, leaving the
two to be inferred independently by their respective channels is necessary.

\textbf{Conditional propagation of factor failure to clustering.} NMI measures only the agreement of the soft membership matrix $D$ with the true partition, and reflects the joint effect of the two channels; a misspecified factor does not necessarily lead to clustering failure. The results of Section~5.1.3 provide a contrast: on networks with a strong topological signal, the NMI of the shared methods is almost unaffected by the factor mismatch---for example, on Adis\_neg of the negative configuration DPSB\_PG still reaches $0.9992$, because here $D$ is determined mainly by the topological channel and the attribute channel only needs not to get in the way; whereas on the A7 series ($\mu=0.7$), where the topological signal is weak, the attribute channel must shoulder community identification independently, and the cost of the factor mismatch is then realized---on the independent instance A7indep, \DCtwo\ gives $0.8708$ whereas DPSB\_PG gives $0.8106$, and on the negative instance A7neg the values are $0.9146$ versus $0.9058$. In other words, the cost of shared modeling is realized precisely in the scenario where attribute information is needed to compensate for an insufficient topology, which is exactly the point of fusing attribute information. 

In real-world networks the ground truths of the two factors are unobservable, but
their observable proxies---the rank correlation between the observed degree $g_i$
and the attribute count $w_i$---are only $-0.04\sim+0.32$ on the four WebKB
sub-networks (Cornell $+0.10$, Texas $+0.32$, Washington $+0.16$, Wisconsin
$-0.04$), far from the break-even point of $+0.89$ (observation noise can only
make this proxy underestimate the true coupling, so the actual coupling cannot be
close to $+0.89$ either). This suggests that in real networks the decoupling of
topological activity and attribute abundance is the norm rather than the
exception, and that the premise on which shared modeling rests is difficult to
satisfy on real data.

\section{Conclusion and outlook}

This paper proposes the Dual-Corrected Stochastic Block Model (\DCtwo). In view of the twofold sources of node heterogeneity in attributed networks, independent correction factors---the topological degree factor $\rho$ and the attribute abundance factor $\eta$---are introduced into the link-generation and the attribute-generation channel, respectively, achieving an effective decoupling of the two kinds of node activity from community membership; a global scaling parameter $\omega$ resolves the mismatch between the multiple normalization constraints and the network size. The model is inferred by the EM algorithm, and the newly introduced factors cancel out completely in the E-step posterior computation, so that no additional computational burden is incurred; the vectorized implementation yields the per-iteration complexity $O(mc^2+nKc+nc\,T_{\text{bis}})$, making the model applicable to large-scale networks. Under specific parameter settings the model reduces to DPSB\_PG or PSB\_PG, thereby inheriting the advantage of the stochastic block model in detecting generalized structures and exhibiting good generality.

The experiments validate the effectiveness of the model on two levels, clustering performance and factor behavior. On 12 synthetic networks and 10 real attributed networks, compared with 6 mainstream methods, \DCtwo\ achieves the highest average NMI on 6 of the 12 synthetic networks and on 5 real networks; compared with the single-shared-factor DPSB\_PG and BCSBM, it is superior on 7 real networks, and its advantage is especially pronounced in the scenario where the topological signal is weak and attribute information must dominate the identification. The factor-level analysis further shows that: (i) the two correction factors can be effectively recovered from the observations of their respective channels (Spearman correlation between $0.64$ and $0.95$); (ii) the two factors are mutually decoupled---when the attribute-side configuration is reversed from perfect positive correlation to perfect negative correlation, the recovery accuracy of the topological factor changes by no more than $0.011$; and (iii) dual independent correction is necessary---the recovery accuracy of single-shared-factor modeling is capped linearly by the true correlation of the two factors, and it is no worse than the independent two-factor scheme only when the two kinds of activity are almost fully synchronized (rank correlation above about $0.9$), whereas in real networks the rank correlation between the observed degree and the attribute count is only $-0.04\sim+0.32$, so that decoupling is the norm rather than the exception. It should be pointed out that the factor-level conclusions rest on the controlled generation assumptions of the synthetic networks, and that the evidence on real networks is confined to the clustering-performance level.

Future work includes: (1) introducing nonparametric Bayesian priors to infer the number of communities automatically; and (2) exploring variational-inference acceleration strategies to accommodate even larger networks.

\begin{acknowledgments}
This work is supported in part by Gansu Provincial University Teacher Innovation Fund Project (2024B-133) and Tianshui Normal University High-Level Institutional Research Project (GJB2024-15). The authors thank the anonymous reviewers for their constructive comments.

\end{acknowledgments}

\section*{\NoCaseChange{CRediT} authorship contribution statement}
\textbf{Zhenhai Chang}: Conceptualization, Methodology, Software, Writing - original draft. \textbf{Shuo Wang}: Software, Validation, Writing - review \& editing. \textbf{Yumeng Liu}: Writing - review \& editing. \textbf{Wei Liu}: Methodology, Writing - review \& editing, Supervision, Project administration.



\appendix

\section{Derivation of the posterior probabilities}
\label{app:posterior}
\renewcommand{\theequation}{A\arabic{equation}}
\renewcommand{\theHequation}{A.\arabic{equation}}
\setcounter{equation}{0}

This appendix gives the derivation of the posterior probability of the edge
latent variable in Section~4.1 (Eq.~\eqref{eq:qij}); the derivation of the
posterior probability of the attribute latent variable (Eq.~\eqref{eq:gamma}) is
entirely analogous.

Assume that each community pair $(r,s)$ independently generates $Z_{ij}^{rs}$
edges following a Poisson distribution with mean
\begin{equation}
\lambda_{ij}^{rs} = \omega\,\rho_i\rho_j\,d_{ir}\theta_{rs}d_{js}.
\end{equation}
By the additivity of the Poisson distribution, for $i<j$ we have
\begin{equation}
A_{ij}=\sum_{r,s}Z_{ij}^{rs} \sim \Poisson\left(\sum_{r,s}\lambda_{ij}^{rs}\right)=\Poisson\left(\omega\rho_i\rho_jB_{ij}\right),
\end{equation}
where $B_{ij}=\sum_{r,s}d_{ir}\theta_{rs}d_{js}$. From the properties of the
conditional Poisson distribution, given the total number of edges $A_{ij}$,
$Z_{ij}^{rs}$ follows a multinomial distribution over all community pairs
$(r,s)$, that is,
$(Z_{ij}^{rs})_{rs}\mid A_{ij},D,\Theta,\rho,\omega \sim \mathrm{Multinomial}(A_{ij},(\frac{\lambda_{ij}^{rs}}{\sum_{r,s}\lambda_{ij}^{rs}})_{rs})$.
Thus, given $A_{ij}$ and the other parameters, the marginal distribution of
$Z_{ij}^{rs}$ over any pair $(r,s)$ is a binomial distribution:
\begin{equation}
Z_{ij}^{rs} \mid A_{ij},D,\Theta,\rho,\omega \sim \mathrm{Binomial}\left(A_{ij},\ \frac{\lambda_{ij}^{rs}}{\sum_{r,s}\lambda_{ij}^{rs}}\right).
\end{equation}
Therefore its posterior expectation is
\begin{equation}
\mathbb E[Z_{ij}^{rs}\mid A_{ij},D,\Theta,\rho,\omega] = A_{ij}\cdot\frac{\lambda_{ij}^{rs}}{\sum_{r,s}\lambda_{ij}^{rs}}
= A_{ij}\cdot\frac{\omega\rho_i\rho_j\,d_{ir}\theta_{rs}d_{js}}{\omega\rho_i\rho_j\sum_{r,s}d_{ir}\theta_{rs}d_{js}}.
\end{equation}
Here $\omega\rho_i\rho_j$ cancels between the numerator and the denominator,
yielding the posterior probability
$q_{ij}^{rs}=d_{ir}\theta_{rs}d_{js}/B_{ij}$ (Eq.~\eqref{eq:qij}); analogously,
on the attribute side, $\gamma_{ik}^r=d_{ir}\phi_{rk}/P_{ik}$
(Eq.~\eqref{eq:gamma}).

\section{Numerical details of the bisection solution of $\lambda_r$ and $\nu$}
\label{app:bisect}
\renewcommand{\theequation}{B\arabic{equation}}
\renewcommand{\theHequation}{B.\arabic{equation}}
\setcounter{equation}{0}

This appendix gives in detail the bisection algorithm used in the M-step to solve
the Lagrange multipliers $\lambda_r$ (Eq.~\eqref{eq:lambda}) and $\nu$
(Eq.~\eqref{eq:eta}). Both equations are univariate monotone functions and can be
solved efficiently by bisection. To prevent division by zero or overflow near the
boundary of the domain, we use a shifting technique to move the root to the
positive half-axis, thereby guaranteeing numerical stability.

\subsection{Solution of $\lambda_r$}

For community $r$, the equation to be solved is
\begin{equation}
f(\lambda) = \sum_{i=1}^n \frac{\rho_i d_{ir} \alpha_{ir}}{\rho_i (C_r + \lambda) + \eta_i} - 1 = 0,
\qquad \lambda > -C_r - \min_i \frac{\eta_i}{\rho_i}.
\label{eq:B1}
\end{equation}
Let $\mathrm{min\_ratio} = \min_{i:\rho_i>0} (\eta_i / \rho_i)$ and set
\[
\mu = \lambda + C_r + \mathrm{min\_ratio},
\]
so that $\mu > 0$ and the denominator can be written as
\[
\rho_i (C_r + \lambda) + \eta_i = \rho_i \mu + b_i,\qquad
b_i = \eta_i - \rho_i \cdot \mathrm{min\_ratio} \ge 0,
\]
with $b_i=0$ for at least one $i$. Transforming Eq.~\eqref{eq:B1} into a function
of $\mu$ gives
\begin{equation}
g(\mu) = \sum_{i=1}^n \frac{\rho_i d_{ir} \alpha_{ir}}{\rho_i \mu + b_i} - 1,
\qquad \mu > 0.
\label{eq:B2}
\end{equation}
It is easy to verify that $g(\mu)$ is continuous and strictly monotonically
decreasing on $(0,+\infty)$, and that
$\lim_{\mu\to 0^+} g(\mu) = +\infty$ (because the denominator of the term with
$b_i=0$ tends to $0$) and $\lim_{\mu\to +\infty} g(\mu) = -1$. Hence the equation
$g(\mu)=0$ has a unique root on the positive half-axis.

\textbf{Bisection steps:}
\begin{enumerate}[1.]
\item If no $\rho_i > \varepsilon$ exists ($\varepsilon=10^{-12}$), return a very
large positive number (e.g., $10^6$) to force the membership of the corresponding
community to zero.
\item Compute $b_i = \eta_i - \rho_i \cdot \mathrm{min\_ratio}$.
\item Find a bracketing interval $[\mu_{\text{lo}}, \mu_{\text{hi}}]$ containing
the root:
  \begin{itemize}
  \item Take the initial values $\mu_{\text{lo}} = \varepsilon$ and
  $\mu_{\text{hi}} = 1.0$.
  \item If $g(\mu_{\text{hi}}) \ge 0$, expand upward by
  $\mu_{\text{hi}} \leftarrow 2\mu_{\text{hi}}$ until $g(\mu_{\text{hi}}) < 0$ or
  the maximum number of expansion steps (40) is reached.
  \item If $g(\mu_{\text{lo}}) \le 0$ (or is infinite), gradually increase
  $\mu_{\text{lo}}$ (e.g., by taking
  $\mu_{\text{lo}} \leftarrow (\mu_{\text{lo}}+\mu_{\text{hi}})/2$) until
  $g(\mu_{\text{lo}}) > 0$. If this is never satisfied, likewise fall back to a
  large positive number.
  \end{itemize}
\item Perform the standard bisection within $[\mu_{\text{lo}}, \mu_{\text{hi}}]$:
  \begin{itemize}
  \item Repeatedly compute $\mu_{\text{mid}} = (\mu_{\text{lo}}+\mu_{\text{hi}})/2$;
  if $g(\mu_{\text{mid}}) > 0$ then $\mu_{\text{lo}} \leftarrow \mu_{\text{mid}}$,
  otherwise $\mu_{\text{hi}} \leftarrow \mu_{\text{mid}}$;
  \item Stop when $|g(\mu_{\text{mid}})| < 10^{-12}$ or the number of iterations
  reaches 80.
  \end{itemize}
\item After solving for $\mu$, recover
$\lambda_r = -C_r - \mathrm{min\_ratio} + \mu$.
\end{enumerate}

In the actual implementation, the denominator of $g(\mu)$ is additionally
protected by $\max(\rho_i \mu + b_i, \varepsilon)$, and extremely large terms are
truncated (e.g., at $10^{15}$) before summation, so that floating-point overflow
is completely avoided. Each evaluation of $g(\mu)$ in the bisection costs $O(n)$,
and the process usually converges within 40--80 iterations; hence the complexity
of solving a single $\lambda_r$ is $O(n)$, and the overall complexity for $c$
communities is $O(nc)$.

\subsection{Solution of $\nu$}

The exact solution of the attribute abundance factor satisfies
\begin{equation}
\eta_i = \frac{w_i}{\sum_r d_{ir} + \nu},\qquad
\sum_{i=1}^n \eta_i = n,
\label{eq:B3}
\end{equation}
where $\nu > -\min_i \big( \sum_r d_{ir} \big)$. Let
$a_i = \max(\sum_r d_{ir}, \varepsilon)$, $a_{\min} = \min_i a_i$, and
$t = \nu + a_{\min}$. Then $t > 0$ and
\[
a_i + \nu = (a_i - a_{\min}) + t = \tilde b_i + t,\qquad
\tilde b_i \ge 0,\;\min_i \tilde b_i = 0.
\]
Define
\begin{equation}
h(t) = \sum_{i=1}^n \frac{w_i}{\tilde b_i + t} - n,
\qquad t > 0.
\label{eq:B4}
\end{equation}
Similarly, $h(t)$ is strictly monotonically decreasing on $(0,+\infty)$,
$\lim_{t\to 0^+} h(t) = +\infty$ and $\lim_{t\to +\infty} h(t) = -n$, so a unique
root exists.

The bisection steps are exactly the same as for $\lambda_r$:
\begin{enumerate}[1.]
\item If all $w_i \le \varepsilon$, return directly $\eta_i = 1$.
\item Take $t_{\text{lo}} = \varepsilon$ and $t_{\text{hi}} = 1.0$ and expand
until $h(t_{\text{hi}}) < 0$; adjust $t_{\text{lo}}$ so that
$h(t_{\text{lo}}) > 0$.
\item Iterate the bisection until convergence (at most 60 iterations).
\item Recover $\nu = t - a_{\min}$, compute
$\eta_i = w_i / (a_i + \nu)$, and normalize to mean $1$.
\end{enumerate}
Each function evaluation costs $O(n)$, and the overall cost is $O(n)$.

\section{Visualization of the correction-factor recovery results}
\label{app:recovery}
\renewcommand{\thefigure}{C.\arabic{figure}}
\renewcommand{\theHfigure}{C.\arabic{figure}}
\setcounter{figure}{0}

\begin{figure}[htbp]
\centering
\includegraphics[width=0.92\linewidth]{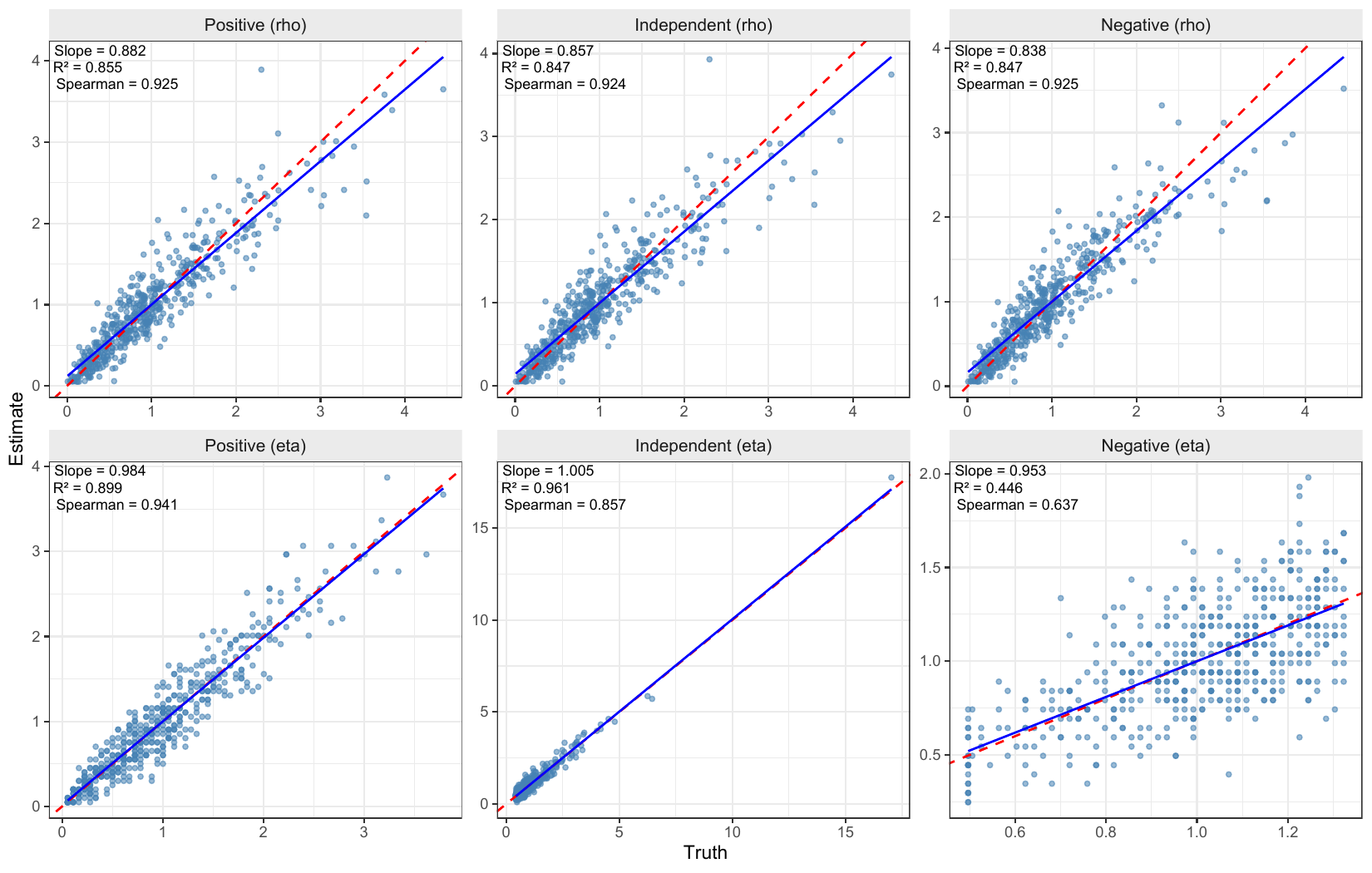}
\caption{Inference of the node-heterogeneity correction factors under the
disassortative structure.}
\label{fig:appC1}
\end{figure}

\begin{figure}[htbp]
\centering
\includegraphics[width=0.92\linewidth]{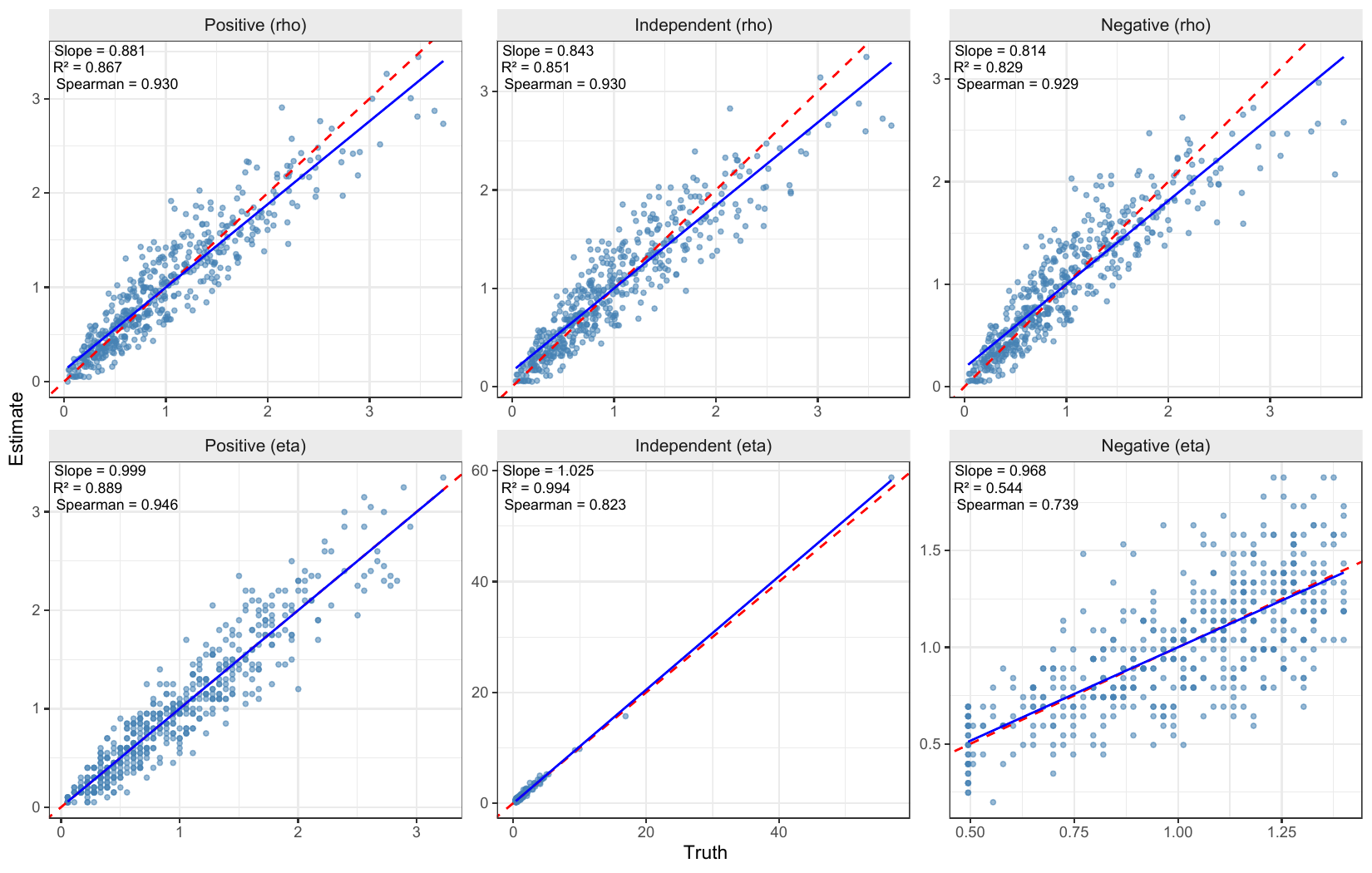}
\caption{Inference of the node-heterogeneity correction factors under the
bipartite structure.}
\label{fig:appC2}
\end{figure}

\end{document}